\documentclass[11pt, a4paper]{article}
\pdfoutput=1

\usepackage{amsmath,color,multicol}
\usepackage{amsfonts}
\usepackage{amssymb}
\usepackage{graphicx}
\usepackage{mathrsfs}
\usepackage{epsfig}
\usepackage{latexsym}
\usepackage{graphicx}
\usepackage{color}
\usepackage{amsmath,amssymb}
\usepackage{cite}
\usepackage{slashed, cancel}
\usepackage{hyperref}
\usepackage{datetime}
\usepackage{cancel}
\usepackage{multirow}
\usepackage{verbatim}
\usepackage{enumitem, array}
\usepackage{multirow}
\usepackage[normalem]{ulem}

\def\circa#1{\,\raise.3ex\hbox{$#1$\kern-.75em\lower1ex\hbox{$\sim$}}\,}

\newcommand{\beq}{\begin{equation}}
\newcommand{\eeq}{\end{equation}}
\newcommand{\beqn}{\begin{eqnarray}}
\newcommand{\eeqn}{\end{eqnarray}}

\newcommand{\bmat}{\begin{pmatrix}}
\newcommand{\emat}{\end{pmatrix}}

\renewcommand{\arraystretch}{1}
\def\app#1#2{%
  \mathrel{%
    \setbox0=\hbox{$#1\sim$}%
    \setbox2=\hbox{%
      \rlap{\hbox{$#1\propto$}}%
      \lower1.1\ht0\box0%
    }%
    \raise0.25\ht2\box2%
  }%
}

\numberwithin{equation}{section}

\font\tenrsfs=rsfs10 at 12pt

\font\sevenrsfs=rsfs7
\font\fiversfs=rsfs5
\newfam\rsfsfam
\textfont\rsfsfam=\tenrsfs
\scriptfont\rsfsfam=\sevenrsfs
\scriptscriptfont\rsfsfam=\fiversfs
\def\mathscr#1{{\fam\rsfsfam\relax#1}}

\hypersetup{colorlinks, citecolor=bluscuro, linkcolor=black, urlcolor=bluscuro}
\definecolor{rossos}{rgb}{0.8,0.2,0.3}
\definecolor{bluscuro}{rgb}{0.15, 0.2, .85}
\definecolor{bluchiaro}{cmyk}{1,.3,0.,0.1}

\makeatother   

 \def\be   {\begin{equation}}   \def\ee   {\end{equation}}
 \def\ba   {\begin{array}}      \def\ea   {\end{array}}
 \def\bea  {\begin{eqnarray}}   \def\eea  {\end{eqnarray}}
 \def\bean {\begin{eqnarray*}}  \def\eean {\end{eqnarray*}}

\begin{document}

%

\begin{flushright} 
OUTP-19-05P
\end{flushright}

\newcommand{\stau}{\widetilde{\tau}}
\newcommand{\bino}{\widetilde{B}}

\vspace{0.5cm}
\begin{center}


{\LARGE \textbf {
A new approach to gauge coupling  unification 
\\[0.3cm]
and proton decay
%
}}
\\ [1.cm]
{\large{\textsc{Stefan Pokorski$^{\,a\,}$, Krzysztof Rolbiecki$^{\,a\,}$, \\ [2mm]
Graham G.~Ross$^{\,b\,}$ and Kazuki Sakurai$^{\,a\,}$
}}}
 \\[0.5cm]

 \textit{
 $^{a}$ Institute of Theoretical Physics, Faculty of Physics,\\University of Warsaw, ~Pasteura 5, PL--02--093 Warsaw, Poland}\\
  \vspace{3mm}

\textit{
$^{b}$ Rudolf Peierls Centre for Theoretical Physics, Clarendon Laboratory\\
University of Oxford, Park Roads, Oxford OX1 3PU, UK.}
 \\  \vspace{3mm}
\end{center}

\vspace{0.7cm}

\begin{center}
\textbf{Abstract}
\begin{quote}

An analytical formalism, including RG running at two loop order, is used to link the supersymmetric and GUT spectra 
in any GUT  model in  which the three gauge couplings unify. 
In each specific GUT model, one can then fully explore the interplay between the pattern of supersymmetry breaking and the prediction for the proton lifetime. 
With this formalism at hand, we study three concrete GUT models:
(i) Minimal $SU(5)$ SUSY GUT, (ii) Missing Partner $SU(5)$ SUSY GUT, and (iii)
an orbifold $SU(5)$ SUSY GUT. In each case we derive interesting conclusions about the possible patterns of the supersymmetric spectrum once the present limits on the proton lifetime are imposed, and vice versa, we obtain  the predictions for the proton lifetime for specific viable choices of the SUSY spectrum.


\end{quote}
\end{center}

\def\thefootnote{\arabic{footnote}}
\setcounter{footnote}{0}
\pagestyle{empty}


\pagestyle{plain}

\newpage
\section{Introduction}
The idea of embedding the  gauge symmetry  groups of the Standard Model (SM) into a larger symmetry, unifying the elementary forces of the SM, is very attractive \cite{Georgi:1974sy}. 
Remarkably, the fermion spectrum of the SM  nicely fits into the multiplets of the simplest Grand Unification symmetry groups,~$SU(5)$ and $SO(10)$.  Moreover, these symmetry groups predict that the SM gauge couplings 
are related to the single underlying GUT gauge coupling for some choice of the superheavy GUT spectrum.
It has been claimed as a big success of supersymmetry (SUSY) that the gauge couplings of the SM
do unify in the Minimal Supersymmetric Standard Model (MSSM), in the presence of SUSY threshold  corrections at ${\cal O}(1)$ TeV and with negligible GUT scale threshold corrections, with the superheavy GUT states having mass at the scale $\sim 10^{16}$ GeV, high enough to sufficiently suppress $D=6$ operator contribution to proton decay. 

However, quantitative analyses in concrete GUT models are much more demanding  and model dependent. Generically, neither the GUT threshold corrections are negligible nor the SUSY spectrum is expected to be degenerate. Constraints from non-observation of proton decay are also model dependent. Among possible decay channels, a special and universal  role is played by the $p \rightarrow \pi^0 e^+$ mode for which the dominant contribution comes from the $D=6$ operators depending almost exclusively on the $X,Y$ gauge boson mass and the value of the unified gauge coupling. In general the $p \to K^+ \nu$ mode, induced by the $D=5$ operators 
generated by the  coloured Higgs exchange diagrams, 
may also give a strong constraint
on the colour triplet Higgs mass and low energy SUSY spectrum 
as well as the structure of the Higgs sector in the GUT models.
However, this mode is highly model-dependent
and several mechanisms have been constructed to eliminate the $D=5$ operator contribution.

The gauge coupling unification implies a non-trivial relation between SUSY and GUT spectra,
which may lead to an interesting interplay between the signatures at collider and proton decay experiments.
A pioneering work has been carried out in Ref.~\cite{Hisano:1992mh,Hisano:1992jj},
where an analytical formula relating SUSY and GUT spectra have been derived at one-loop level in the minimal SU(5) SUSY GUT model,
with two-loop running included numerically.
Similar one-loop formulae have also been presented for the models proposed in \cite{Hisano:1994fn} and \cite{Hall:2001pg}.
The interplay  of the SUSY and GUT threshold corrections in the
unification of the gauge couplings in the MSSM has been studied in various contexts, mostly by numerical methods 
\cite{Ross:1978wt,
Marciano:1979yg,
Goldman:1980ah,
Ellis:1980jm,
Murayama:2001ur,
Maekawa:2002mx,
Acharya:2008zi,
Hisano:2013exa,
Hisano:2013cqa,
Maekawa:2013sua,
Maekawa:2014gva,
Hebecker:2014uaa,
Wang:2015mea,
Ellis:2015jwa,Bajc:2016qcc}. 
Most such works have focused on constraining the GUT spectra, assuming the low energy supersymmetric spectrum is degenerate at $m_{SUSY}$. 
In this paper we  propose to use an analytical formalism linking the SUSY and GUT spectra
based on the parametrization of the one-loop SUSY and GUT threshold corrections in terms of three effective parameters describing each of them.\footnote{
  Similar parametrizations of SUSY spectra have been previously used in Refs.~\cite{Langacker:1992rq,Carena:1993ag,Krippendorf:2013dqa}. }
The link between the two spectra necessary to ensure the gauge coupling unification at two-loop level
can then be expressed as two relations between two pairs of these parameters.
The effective parameters are calculable in terms of particle masses.  The formalism provides a convenient way to analyse 
general GUT models with arbitrary SUSY spectra.
In each specific GUT model, one can then fully explore the interplay between the pattern of supersymmetry breaking and the prediction for the proton lifetime.
We apply this formalism to four concrete examples of the GUT models.  In each case we derive interesting conclusions about the possible patterns of the supersymmetric spectrum once the present limits on the proton lifetime are imposed, and vice versa, we obtain  the predictions for the proton lifetime for specific viable choices of the SUSY spectrum.

\section{The formalism}

In this paper we assume that there is an underlying simple GUT group unifying the SM gauge couplings.
The relation of the gauge couplings at $m_Z$ to the universal gauge coupling evaluated at a scale  $\Lambda$, of order the superheavy masses,
is given by solving the renormalization group equations (RGEs). 
The solution can be written as \cite{Weinberg:1980wa, LLewellynSmith:1980mu, Hall:1980kf, Langacker:1992rq}
\begin{equation}
\frac{2 \pi}{\alpha(\Lambda)} \,=\, \frac{2 \pi}{\alpha_i(m_Z)} - 
b_i \, \ln \frac{\Lambda}{m_Z} + 
s_i + r_i + \gamma_i + \Delta_i \,,
\label{eq:RGE}
\end{equation}
where 
$\alpha_1 \equiv \frac{5}{3} \alpha_Y$,
$i = 1,2,3$, represents the gauge group  and
$b_i = (\frac{33}{5}, 1, -3)$ are the one-loop $\beta$-function coefficients for the MSSM. 
The quantities $s_i$ and $r_i$ represent the low energy supersymmetric and GUT scale threshold corrections, respectively. 
(See the Appendix for the derivation and interpretations of the threshold corrections.)
They read
\beq
s_i \, = \, \sum_\eta b_i^\eta \ln \frac{m_\eta}{m_Z} \,,
\label{eq:s}
\eeq
and 
\beq
r_i \,=\, \sum_\xi b_i^\xi \ln \frac{m_\xi}{\Lambda}\,.
\label{eq:r}
\eeq 
The parameters 
$m_\eta$ and $b_i^\eta$
denote the mass and the contribution to $b_i$
from the superparticle $\eta$. 
The parameters $m_\xi$ and $b_i^\xi$ are the corresponding parameters for the  GUT scale particle $\xi$. Both spectra are arbitrary at this point.  
Clearly, gauge coupling unification puts strong constraints on the sums $s_i+r_i$ that we are going to quantify in the following.
%
The $\gamma_i \equiv - \frac{1}{2} \sum_i \frac{b_{ij}}{b_i} \ln \big( \frac{\alpha_j(\Lambda)}{\alpha_j(m_Z)} \big)$ 
accounts for the two-loop contribution
with $b_{ij}$ being the two-loop $\beta$-function coefficients.\footnote{ At the scale $\Lambda$,
it can be approximated as $\gamma_i = \frac{1}{4 \pi} \sum_i \frac{b_{ij}}{b_i} \ln \big( 1 + \frac{b_j \alpha_\Lambda}{2 \pi} \ln \frac{\Lambda}{m_Z} \big)$.
One can solve $\gamma_i$ iteratively by updating $\Lambda$ and $\alpha_\Lambda$ \cite{Langacker:1992rq}.}
The $\Delta_i$ represents the effect of the top Yukawa coupling
and the conversion factor between $\overline{MS}$ and $\overline{DR}$ schemes.\footnote{For the treatment of $\Delta_i$, see for example \cite{Langacker:1992rq}.}
%


Since any 3-dimensional vector can be expanded in terms of 3 independent vectors, 
the three terms in Eq.~\eqref{eq:RGE}, $[(2 \pi \alpha^{-1}(m_Z) + \gamma + \Delta)_i$, $s_i$, $r_i$], can be expanded 
in terms of [(1,1,1), $b_i$, $\delta_i$] as
\beqn
\frac{2 \pi}{\alpha_i(m_Z)} &=& \frac{2 \pi}{\alpha_G^*} + b_i \ln \frac{M_G^*}{m_Z}  - \delta_i \ln \frac{M_S^*}{m_Z} - \gamma_i - \Delta_i \,,
\nonumber \\
s_i &=& C_S + b_i \ln \Omega_S + \delta_i \ln \frac{T_S}{m_Z} \,,
\nonumber \\
r_i &=&C_G - b_i \ln \frac{T_G}{\Lambda} - \delta_i \ln \Omega_G \,,
\label{eq:ass}
\eeqn
where $b_i$ is the MSSM $\beta$-function coefficients
and $\delta_i \equiv b_i - b_i^{\rm SM}$ is the difference between those and the SM ones.
The parameters $C_S, T_S, \Omega_S$ and $C_G, T_G, \Omega_G$ fully parametrize any arbitrary supersymmetric and GUT threshold corrections 
at the leading logarithmic level. 
They can be found by solving the second and third set of the equations, once $s_i$ and $r_i$ are given in terms of concrete spectra.

The first set of the above equations can be interpreted as the solution to the RGE for the special case where
the GUT threshold correction is absent and all SUSY particles are degenerate at $M_S^*$.
The $M_G^*$ and $\alpha_G^{*-1}$ are then the unification scale and the unified coupling for this idealised situation, respectively.
We find numerically that $\alpha_G^{*-1} = 25.5$, $M_G^* = 1.26 \cdot 10^{16}$ GeV, $M_S^* = 2.13$ TeV solve the first set of equations for the experimental values of the gauge couplings at $m_Z$ with the recent world average $\alpha_s^0(m_Z) = 0.1183$ \cite{dEnterria:2018cye}.
By varying the $\alpha_s(m_Z)$ within its 1-$\sigma$ error, $\Delta \alpha_s = 0.0008$ \cite{dEnterria:2018cye}, 
we have $M_S^* \in [2.69, 1.72]$\,TeV, $M_G^* \in [1.17, 1.35]\cdot 10^{16}$\,GeV
and $\alpha_G^* \in [25.7, 25.4]$, where the left and right values correspond to the 
negative and positive variation of the strong coupling $\alpha_s(m_Z) \in [0.1175, 0.1191]$.
For more general cases, those constants can be approximately written in terms of $\alpha_s(m_Z)$ as
\beqn
\frac{M_S^*}{\rm TeV} &=& \frac{2.13}{\rm TeV} \cdot \exp \left[ - 0.224 \left( \frac{\alpha_s - \alpha_s^0}{\Delta \alpha_s} \right) \right], \nonumber \\
\frac{M_G^*}{\rm GeV} &=& \frac{1.26 \cdot 10^{16}}{\rm GeV} \cdot \exp \left[ 0.0715 \left( \frac{\alpha_s - \alpha_s^0}{\Delta \alpha_s} \right) \right],
\nonumber \\
\alpha_G^{*-1} &=& 25.5 - 0.172 \left( \frac{\alpha_s - \alpha_s^0}{\Delta \alpha_s} \right)\,.
\label{eq:as_effect}
\eeqn



It is convenient to trade the three experimental numbers, $\alpha_i(m_Z)$, for the three new parameters $\alpha_G^*, M_G^*, M_S^*$.
Substituting Eq.~\eqref{eq:ass} into Eq.~\eqref{eq:RGE}, we get
\beqn
\frac{2 \pi}{\alpha(\Lambda)} 
\,=\,
\Big[ \frac{2 \pi}{\alpha_G^*} + C_S + C_G \Big]
+ b_i \ln \Big( \frac{M_G^* \Omega_S}{T_G} \Big) 
+ \delta_i \ln \Big( \frac{T_S}{M_S^* \Omega_G} \Big) \,.
\eeqn
Since the three basis-vectors of this expansion are independent, 
in order for the left-hand-side to be $i$-independent the two logarithms in the right-hand-side must vanish.  
This is the condition of the gauge coupling unification.
Namely, the GUT and SUSY spectra must satisfy the following simultaneous conditions:
\\
\beq
\boxed{
~~~T_S ~=~ M_S^* \Omega_G~~~\cap~~~T_G ~=~ M_G^* \Omega_S~~~
\label{eq:GCUcond}
}
\eeq
\vspace{-1mm}
\\
The inverse of unified gauge coupling is then given by:
\\
\beq
\boxed{
~~~
\alpha^{-1}(\Lambda) ~=~ \alpha_G^{*-1} + \frac{C_S + C_G}{2 \pi} 
~~~
\label{eq:unifiedcoupling}
}
\eeq
\vspace{-1mm}
\\

The second and third equations in Eq.~\eqref{eq:ass} can easily be solved.
The general solutions can be written as
\beqn
\ln \frac{T_S}{m_Z} &=& v_i s_i / D \,,
\nonumber \\
\ln \Omega_S &=& u_i s_i / D \,,
\nonumber \\
C_S &=& -\epsilon_{ijk} b_i \delta_j s_k / D \,,
\nonumber \\
\ln \Omega_G &=& - v_i r_i / D \,,
\nonumber \\
\ln \frac{T_G}{\Lambda} &=& - u_i r_i / D \,,
\nonumber \\
C_G &=& -\epsilon_{ijk} b_i \delta_j r_k/ D \,,
\eeqn
with
\beqn
&&
v = 
\bmat
b_2 - b_3 \\
-b_1 + b_3 \\
b_1 - b_2
\emat
=
\bmat
4 \\ - \frac{48}{5} \\ \frac{28}{5}
\emat \,,
~~~~~~
u = 
\bmat
-\delta_2 + \delta_3 \\
\delta_1 - \delta_3 \\
-\delta_1 + \delta_2
\emat
=
\bmat
-\frac{1}{6} \\ - \frac{3}{2} \\ \frac{5}{3}
\emat,
\nonumber
\eeqn
\beq
D ~=~ b_2 \delta_1 - b_3 \delta_1 - b_1 \delta_2 + b_3 \delta_2 + 
 b_1 \delta_3 - b_2 \delta_3
 = - \frac{38}{5} \,,
\eeq
where we used $b_i = (\frac{33}{5}, 1, -3)$, $b^{\rm SM}_i = (\frac{41}{10}, -\frac{19}{6}, -7)$
and $\delta_i = b_i - b_i^{\rm SM} = (\frac{5}{2}, \frac{25}{6}, 4)$.
Substituting these numbers the solutions become \cite{Pokorski:2017ueo}
\beqn
T_S &=& \Big[ M_3^{-28} M_2^{32}  \mu^{12}
 m_A^{3}  X_{T} \Big]^{\frac{1}{19}},
 \nonumber \\
\Omega_S &=& \Big[ M_3^{-100} M_2^{60} \mu^{32} m_A^{8} X_{\Omega} \Big]^{\frac{1}{288}}\,,
\nonumber \\
C_S &=& \frac{125}{19} \ln M_3 - \frac{113}{19} \ln M_2 - \frac{40}{19} \ln \mu - \frac{10}{19} \ln m_A \nonumber \\
&+& \sum_{i=1...3} \Big[ 
 \frac{79}{114} \ln m_{\tilde d_{Ri}}
- \frac{10}{19} \ln m_{\tilde l_i}
- \frac{121}{114} \ln m_{\tilde q_i} 
+ \frac{257}{228} \ln m_{\tilde u_{Ri}}
+ \frac{33}{76} \ln m_{\tilde e_{Ri}}
\Big]\,,
\label{eq:formula_s}
\eeqn
with
\beqn
X_{T} \,\equiv\, \prod_{i=1...3} \Big( \frac{m^3_{\tilde l_i}}{m^3_{\tilde d_{Ri}}} \Big) \Big( \frac{m^7_{\tilde q_i}}{m^2_{\tilde e_{Ri}} m^5_{\tilde u_{Ri}}}  \Big)\,,
~~~~~
X_{\Omega} \,\equiv\, \prod_{i=1...3} \Big( \frac{m^8_{\tilde l_i}}{m^8_{\tilde d_{Ri}}} \Big)
\Big( \frac{m^6_{\tilde q_i}m_{\tilde e_{Ri}}}{m^7_{\tilde u_{Ri}}} \Big)\,,
\eeqn
for the MSSM sparticles and
\beqn
\ln \Omega_G &=& \sum_\xi \Big( \frac{10}{19} b_1^\xi - \frac{24}{19} b_2^\xi + \frac{14}{19} b_3^\xi \Big) \ln \frac{m_\xi}{\Lambda} \,,
\nonumber \\
\ln \frac{T_G}{\Lambda}  &=& \sum_\xi \Big( -\frac{5}{228} b_1^\xi - \frac{15}{76} b_2^\xi + \frac{25}{114} b_3^\xi \Big) \ln \frac{m_\xi}{\Lambda}  \,,
\nonumber \\
C_G &=& 
\sum_\xi \Big( \frac{165}{76} b_1^\xi - \frac{339}{76} b_2^\xi + \frac{125}{38} b_3^\xi \Big) \ln \frac{m_\xi}{\Lambda} \,,
\label{eq:formula_g}
\eeqn
for the GUT scale particles. In Eq.~\eqref{eq:formula_s} the parameters $M_2$, $M_3$, $m_A$ and $\mu$ denote the wino, gluino, 
the CP-odd scalar soft masses and the higgsino mass, respectively, each at the corresponding decoupling (threshold) scale. 

The condition Eq.~\eqref{eq:GCUcond} with Eqs.~\eqref{eq:formula_s} and \eqref{eq:formula_g} 
have to be satisfied for arbitrary supersymmetric and GUT scale spectra to ensure the gauge coupling unification. 
These compact relations can be used, for instance, to discuss  the patterns of the supersymmetric spectra consistent with the unification in various concrete GUT models; e.g.~conventional models in 4d or in extra dimensional  models, with the present and future limits on the proton decay imposed. They quantify a non-trivial interconnection between collider and proton decay experiments.  In the following we shall discuss the implications of the gauge coupling unification, i.e.~Eq.~\eqref{eq:GCUcond},
in several concrete GUT models.

\section{Minimal $SU(5)$}
\label{sec:msu5}

The first example in which we apply our formula and study 
an interplay between the low energy SUSY and GUT spectra is the minimal $SU(5)$ model. 
The Higgs sector of this model contains 
the adjoint chiral multiplet $\Sigma({\bf 24}) = (\Sigma_8, \Sigma_{(2,3)}, \Sigma_{(2, \bar 3)}, \Sigma_3, \Sigma_1)$ and 
the (anti-)fundamental chiral multiplet $H({\bf 5}) = (H_C, H_u)$ ($\overline H(\overline {\bf 5}) = (\overline H_C, \epsilon H_d)$). 
In the above notation, $\Sigma_i$ are the component of $\Sigma$ under $G_{\rm SM} = U(1) \times SU(2) \times SU(3)$ decomposition, 
$H_C$ ($\overline H_C$) is the colour (anti-)triplet Higgs field
and $H_{u/d}$ are the doublet Higgs fields in the MSSM. 
The Higgs superpotential is given by
\beq
W_{H} \,=\, \frac{1}{2} M {\rm Tr} \Sigma^2 + \frac{1}{3} \lambda_\Sigma {\rm Tr} \Sigma^3  + 
\lambda_H \overline H (\Sigma + 3 V) \Sigma H, 
\label{eq:WH5}
\eeq
where the dimensionfull parameter $V$ is related to the VEV of $\Sigma$ as
\beq
\langle \Sigma \rangle \,=\, V \cdot {\rm diag}(2,2,2,-3,-3) \,,
\label{eq:vev5}
\eeq
such that the MSSM Higgs fields become massless by the cancellation in the last term of Eq.~\eqref{eq:WH5}.
This cancellation is called doublet-triplet splitting problem since it requires an enormous fine-tuning.
The mass of colour triplet Higgses reads
\beq
M_{H_C} \,=\, 5 \lambda_H V\,.
\eeq 
The direction of the VEV in Eq.~\eqref{eq:vev5} breaks the $SU(5)$ gauge symmetry to $G_{\rm SM}$,
giving masses to the $X, Y$ gauge bosons
\beq
M_{V} \,=\, 5 \sqrt{2} g_5 V \,,
\eeq
with $g_5$ being the $SU(5)$ gauge coupling.
While the Goldstone components $\Sigma_{(3,2)}, \Sigma_{(\bar 3,2)}$ in the $\Sigma$ field are eaten by the $X,Y$ fields,
the $\Sigma_8$ and $\Sigma_3$ components have the mass
\beq
M_\Sigma \,=\, \frac{5}{2} M \,,
\eeq
and contribute to the GUT threshold correction together with the $X,Y$ and $H_C, \overline H_C$ fields.
The singlet field $\Sigma_1$ has the mass $\frac{1}{2} M$ but does not contribute to the
GUT threshold correction, thus is irrelevant to our discussion.

\begin{table}[t!]
\begin{center}
\begin{tabular}{|c|c|c|}
\hline
~~~mass~~~ & $(U(1) \times $SU(2)$ \times $SU(3)$)$ & $(b_1, b_2, b_3)$ \\
\hline
$M_{H_C}$ & $(-\frac{1}{3}, {\bf 1}, {\bf 3}),~ (\frac{1}{3}, {\bf 1}, {\bf \overline 3})$ & $(\frac{2}{5}, 0, 1)$ 
\\ \hline
$M_V$ & $(-\frac{5}{6}, {\bf 2}, {\bf 3}),~ (\frac{5}{6}, {\bf 2}, {\bf \overline 3})$ & $(-10, -6, -4)$ 
\\ \hline
$M_\Sigma$ & $(0, {\bf 3}, {\bf 1}),~ (0, {\bf 1}, {\bf 8})$ & $(0, 2, 3)$ 
\\ \hline
\end{tabular}
\caption{\small The GUT mass spectrum and the contribution to the $\beta$-function coefficients in the Minimal $SU(5)$ model.
    \label{tb:su5}
}
\end{center}
\end{table}

We summarise the field components that contribute to the GUT threshold correction in Table~\ref{tb:su5}.
By plugging the masses and $\beta$-function coefficients into the formula \eqref{eq:formula_g}, we find
\beqn
\Omega_G &=& M_{H_C}^{\frac{18}{19}} M_V^{-\frac{12}{19}} M_\Sigma^{-\frac{6}{19}} \,,
\nonumber \\
T_G &=& M_{H_C}^{\frac{4}{19}} M_V^{\frac{10}{19}} M_\Sigma^{\frac{5}{19}} \,,
\nonumber \\
C_G &=& 
3 \ln \frac{\Lambda}{M_G^*} 
+ \frac{79}{19} \ln  \frac{M_{H_C}}{M_G^*}  
- \frac{154}{19} \ln  \frac{M_V}{M_G^*}  
+ \frac{18}{19} \ln  \frac{M_\Sigma}{M_G^*}  \,,
\label{eq:OTC}
\eeqn
where we have used $b^G = (-3,-3,-3)$ for the full Minimal $SU(5)$ $\beta$-function coefficients.
Using the unification conditions Eq.~\eqref{eq:GCUcond}, one obtains
\beqn
M_{H_C} &=&  M_G^* \Omega_S \left( \frac{T_S}{M_S^*} \right)^{\frac{5}{6}} \,,
\label{eq:MHCsu5}
\\
(M_V^2M_\Sigma)^{\frac{1}{3}} &=& M_G^* \Omega_S \left( \frac{T_S}{M^*_s} \right)^{-\frac{2}{9}}
\label{eq:MASSsu5} \,.
\eeqn
It is worth noting that despite the look of Eq.~\eqref{eq:formula_g}, these conditions 
do not depend on $\Lambda$. A general proof of the $\Lambda$ independence and an exceptional case 
are given and discussed in the  Appendix.
The above equations are remarkable in the sense that they allow us to analytically calculate 
the masses of superheavy particles in terms of the low energy SUSY spectrum 
through $T_S$ and $\Omega_S$ given in Eq.~\eqref{eq:formula_s}.

\begin{figure}[!t!]
\begin{center}
\includegraphics[width=5.6cm]{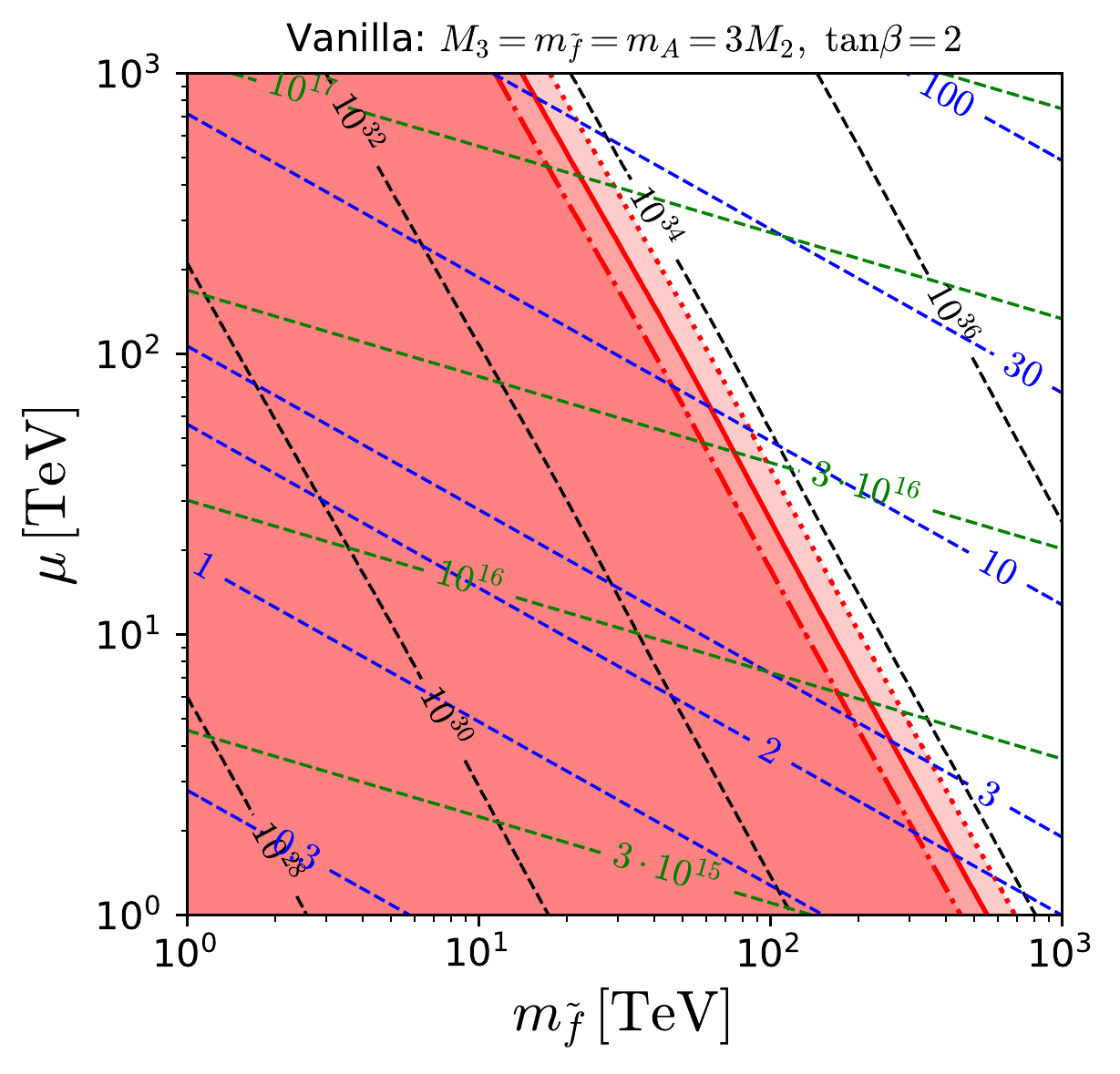}
\includegraphics[width=5.6cm]{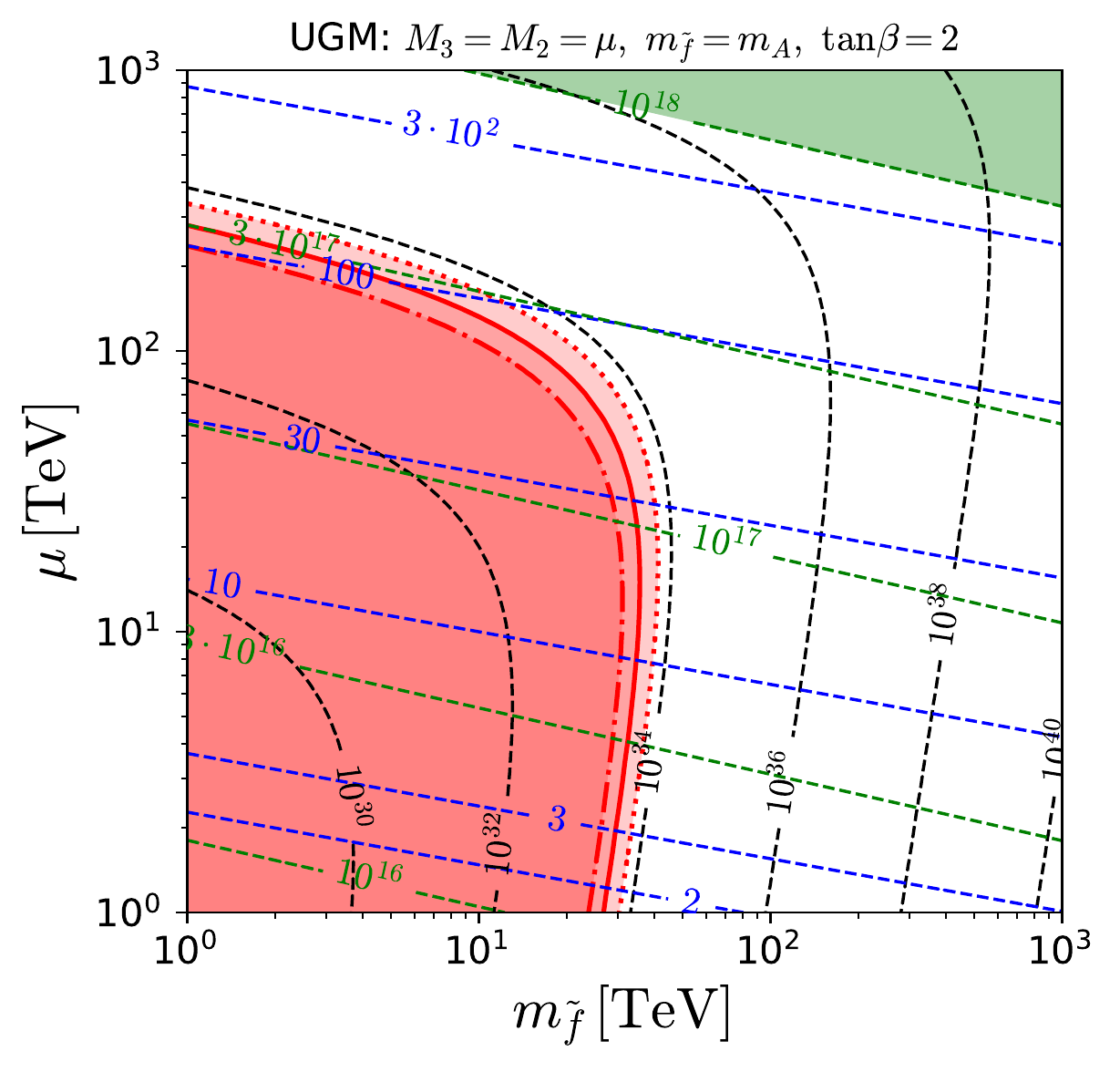}
\includegraphics[width=5.6cm]{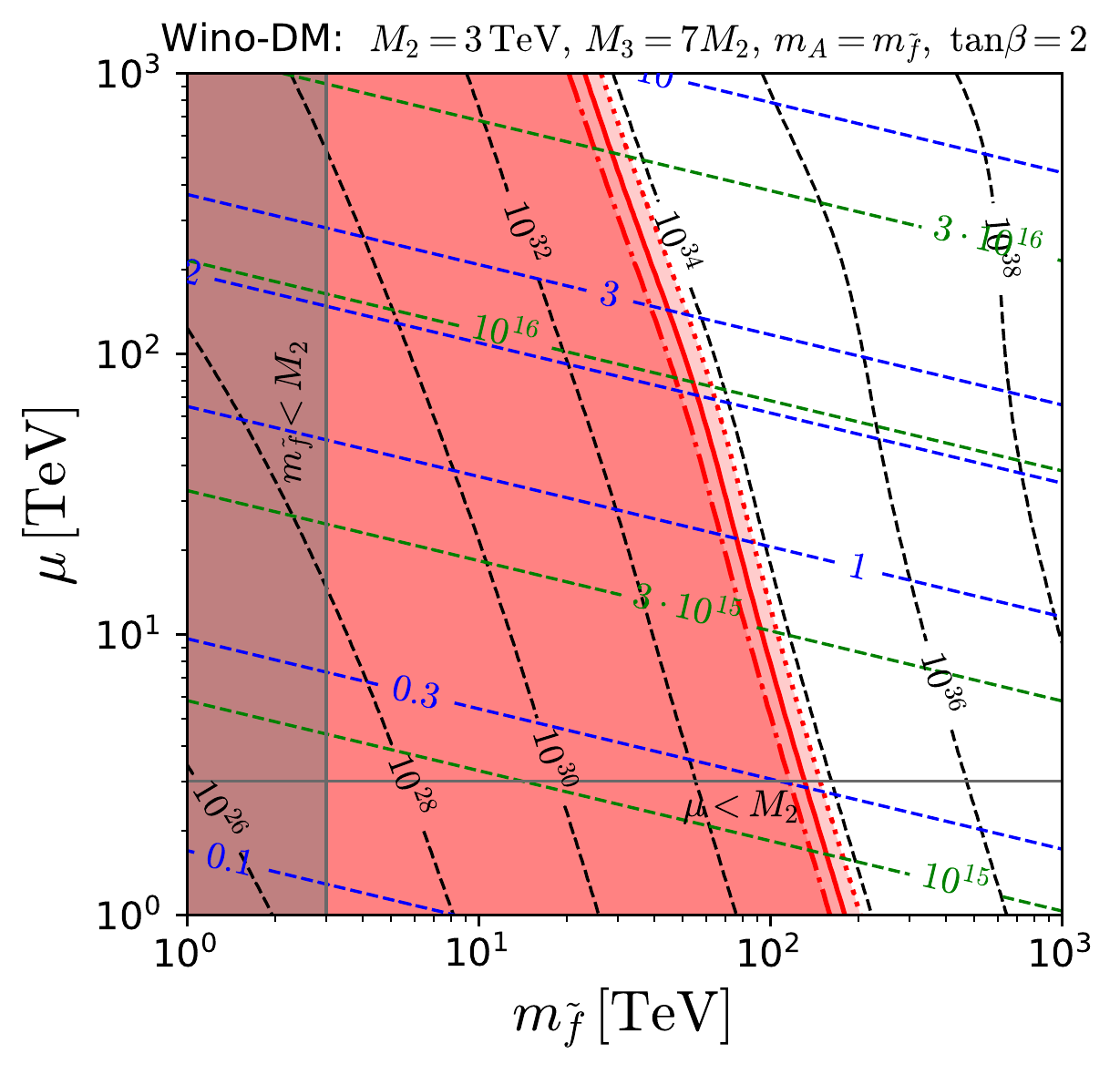}
\vspace{-2mm}
\caption{\label{fig:su5}
\small The low energy SUSY parameter space ($m_{\tilde f}$ versus $\mu$ plane) for three different scenarios:
{\bf Left}; ``vanilla'' SUSY with $M_3 = 3 M_2 = m_A = m_{\tilde f}$ and $\tan\beta = 2$. 
{\bf Centre}; UGM scenario with $M_3 = M_2 = \mu, m_A = m_{\tilde f}$ and $\tan\beta = 2$.
{\bf Right}; Wino DM scenario with $M_2 = 3$ TeV, $M_3 = 7 M_2$, $m_A = m_{\tilde f}$ and $\tan\beta = 2$.
The light red region is excluded due to the current limit, $\tau(p \to K^+ \bar \nu) > 4.0 \cdot 10^{33}$ yrs.
The band around the boundary of the red region represents the uncertainty coming from $\alpha_s(m_Z)$.
The dashed-dotted and dotted contours correspond to the limit obtained from $\alpha_s(m_Z) = \alpha_s^0 \pm \Delta \alpha_s$, respectively.
The dashed black, green and blue contours represent the values of $\tau(p \to K^+ \bar \nu)$/yrs,
$M_{H_C}$/GeV and $T_S$/TeV, respectively.
The shaded green region is disfavoured because $M_{H_C}$ is close to the Planck scale.
}
\end{center}
\end{figure}

Eq.~\eqref{eq:MHCsu5} is particularly interesting since
it allows the prediction of the $D=5$ proton decay rate purely from the low energy SUSY spectrum.\footnote{In the calculation of the $D=5$ proton decay, we closely follow Ref.~\cite{Hisano:2013exa}.
We thank J.~Hisano and N.~Nagata for the details of the calculation. }
The results are plotted in Fig.~\ref{fig:su5} in the  ($m_{\tilde f}$, $\mu$) SUSY plane, where $m_{\tilde f}$ is the universal sfermion mass, for three scenarios: 
(Left; ``Vanilla'' SUSY,~ Centre; Universal Gaugino Mass (UGM) scenario,~ Right; Wino DM scenario. The details of these scenarios are given at the end of this Section.)

In the plots, the light red region is excluded by the current proton decay limit 
$\tau(p \to K^+ \bar \nu) > 4.0 \cdot 10^{33}$ yrs.
The band around the boundary of the red region represents the uncertainty coming from $\alpha_s(m_Z)$.
The dashed-dotted and dotted contours correspond to the limit obtained from 
the upper and lower 1-$\sigma$ variations of $\alpha_s(m_Z)$, respectively.
The dashed black, green and blue contours represent the values of $\tau(p \to K^+ \bar \nu)$/yrs,
$M_{H_C}$/GeV and $T_S$/TeV, respectively.
The shaded green region is disfavoured because there $M_{H_C}$ is very close to the Planck scale.

We also compute $(M_V^2 M_\Sigma)^{\frac{1}{3}}$ and 
the unified coupling $\alpha^{-1}(\Lambda)$
for $\Lambda = {\rm max}\{M_{H_C}, M_V\}$,
assuming $M_V = M_\Sigma$,
for our three example scenarios.
We found that $(M_V^2 M_\Sigma)^{\frac{1}{3}}$ varies very mildly due to the small power in Eq.~\eqref{eq:MASSsu5}. Over the region shown on the plots we find:
 \beqn
{\rm Vanilla\,\,SUSY}:&& 3.77 \cdot 10^{15}\,{\rm GeV} < (M_V^2 M_\Sigma)^{\frac{1}{3}} < 1.83 \cdot 10^{16}\,{\rm GeV} 
\nonumber \\
{\rm UGM}:&& 3.14 \cdot 10^{15}\,{\rm GeV} < (M_V^2 M_\Sigma)^{\frac{1}{3}} < 1.52 \cdot 10^{16}\,{\rm GeV} 
\nonumber \\
{\rm Wino\,\,DM}:&& 8.90 \cdot 10^{15}\,{\rm GeV} < (M_V^2 M_\Sigma)^{\frac{1}{3}} < 1.19 \cdot 10^{16}\,{\rm GeV} 
\nonumber 
\eeqn 
Generally, light $X,Y$ gauge bosons with $M_V \lesssim 10^{16}$ GeV may induce observably large $D=6$ proton decay
($p \to \pi^0 e^+$).
However, within the above range, the $D=6$ proton decay constraint can always be avoided by
lowering $M_\Sigma$, which does not lead to any phenomenologicallty dangerous processes. 
For completeness we also report the range of the unified coupling under the assumption $M_V = M_\Sigma$.
We found in the region of the plots:
\beqn
{\rm Vanilla\,\,SUSY}:&& 24.7 < \alpha^{-1}(\Lambda) < 31.8
\nonumber \\
{\rm UGM}:&& 25.0 < \alpha^{-1}(\Lambda) < 33.0
\nonumber \\
{\rm Wino\,\,DM}:&& 25.3 < \alpha^{-1}(\Lambda) < 29.5
\nonumber 
\eeqn 
As can be seen, $\alpha^{-1}(\Lambda)$ is always in a perturbative regime within the region of interest.

The ``Vanilla" SUSY scenario, shown in the left panel of Fig.~\ref{fig:su5}, assumes 
 a low energy gaugino mass ratio $M_3/M_2 = 3$.
This ratio typically arises 
in the scenarios where the gaugino masses are unified around the GUT scale,
e.g.~in the Constrained MSSM (CMSSM) or in the Gauge Mediated SUSY Breaking (GMSB) scenario.
For simplicity we also assume all sfermion masses are degenerate at their mass scale, $m_{\tilde f}$,
and $M_3 = m_A = m_{\tilde f}$.
We take $\tan\beta = 2$ for all three scenarios in Fig.~\ref{fig:su5}.
For larger values of $\tan\beta$, the proton decay constraint is more constraining since 
the $D=5$ proton decay rate grows with some positive powers of $\tan \beta$.\footnote{ 
The amplitude of the $D=5$ proton decay scales as $\tan^2 \beta$ for the Higgsino exchange diagram
and $\tan \beta$ for the Wino exchange diagram.
In the most region of our numerical scan, the contribution from Wino exchange diagram dominates the decay rate.}
In the left plot, we see that in the Vanilla SUSY scenario the proton decay constraint requires 
$M_3 > 500$ TeV for $\mu \sim 1$ TeV and $M_3 > 20$ TeV for $\mu \sim 10^3$ TeV.

The Universal Gaugino Mass (UGM) scenario shown in the central panel of Fig.~\ref{fig:su5}, assumes the  low energy gaugino mass ratio $M_2/M_3 = 1$.
Such a gaugino mass ratio may arise in the gaugino focus point scenario \cite{Abe:2007kf,Horton:2009ed,Kaminska:2013mya}.
For other parameters we take $M_3 =\mu$ and $m_A = m_{\tilde f}$ as an example.
We see that light SUSY spectra are allowed by the proton decay limit apart from the sfermion masses, which have $m_{\tilde f} > 30$ TeV.

The ``Wino DM" SUSY scenario shown in the right panel of Fig.~\ref{fig:su5} has the dark matter abundance dominated by the Wino.
It has been shown that the thermal Wino abundance with  $M_{\widetilde W}\sim 3$ TeV can account for the observed energy density 
of the DM in the present Universe. 
Since the Wino becomes the lightest gaugino in the Anomaly Mediated SUSY Breaking (AMSB) scenario,
we assume the low energy gaugino mass relation $|M_3/M_2| = 7$ predicted by the AMSB.
For simplicity, we further assume a universal sfermion mass and $m_A = m_{\tilde f}$.
In the plot, the shaded grey region is disfavoured since the sfermions are lighter than the Wino.
Below the horizontal grey line, $|\mu| < M_2$ but we take this region into our consideration
since the thermal Higgsino may account for the DM relic density in this region.
We see in this plot that the current proton decay limit demands
the sfermions to be heavier than 200 TeV for $\mu \sim 1$ TeV and heavier than 40 TeV for $\mu \sim 10^3$ TeV.  
This means that the split SUSY scenarios with both Wino and Higgsino DM are consistent with
the minimal $SU(5)$ model.  
In particular, many such scenarios predict loop suppressed gaugino masses compared to the sfermion mass,
$m_{\lambda}/m_{\tilde f} \sim 1/100$. 
Thus Wino or Higgsino DM models in minimal $SU(5)$ predict a proton decay lifetime in the region that may be discovered by  the next generation experiments.

\section{Missing partner $SU(5)$ models}
\label{sec:missing_su5}

\subsection{Hagiwara-Yamada Model}

We now study a model presented in Ref.~\cite{Hagiwara:1992ys}.
The field content in the Higgs sector is given in Table \ref{tb:mpsu5}.
\begin{table}[h!]
\begin{center}
\begin{tabular}{|c||c|c|c|c|c|c|c|c|c|c|}
\hline
Field     & $H$ & $\overline H$ & $\Theta$ & $\overline \Theta$ & $\Sigma$ & $X$  \\
\hline
rep.      & $\bf 5$ & $\bf \overline 5$ & $\bf 50$ & $\bf \overline{50}$ & $\bf 75$ & $\bf 1$ \\
\hline
$U(1)_{X}$ & $-2$ & $1$ & $2$ & $-1$ & $0$ & $-1$ \\
\hline
\end{tabular}
\caption{The field content in the Higgs sector.
    \label{tb:mpsu5}
}
\end{center}
\end{table}
The superpotential of the Higgs sector is given by
\beq
W_H \,=\, W_1 + W_2 \,.
\label{eq:WH}
\eeq
The first term is the superpotential containing only $\Sigma$:
\beq
W_1 \,=\, M_\Sigma \Sigma^2 - \frac{1}{3} \lambda_{75} \Sigma^3,
\eeq
which let $\Sigma$ develop a VEV that breaks $SU(5)$ into $G_{\rm SM}$.
We have
\beqn
\langle \Sigma \rangle^{[\alpha \beta]}_{[\gamma \delta]} &=& \frac{1}{2} \left\{ \delta^\alpha_\gamma \delta^\beta_\delta - \delta^\alpha_\delta \delta^\beta_\gamma \right\} V_\Sigma \,,
\nonumber \\
\langle \Sigma \rangle^{[a b]}_{[c d]} &=& \frac{3}{2} \left\{ \delta^a_c \delta^b_d - \delta^a_d \delta^b_c \right\} V_\Sigma \,,
\nonumber \\
\langle \Sigma \rangle^{[a \alpha]}_{[b \beta]} &=& -\frac{1}{2} \left\{ \delta^a_b \delta^\alpha_\beta \right\} V_\Sigma \,,
\eeqn  
with 
\beq
V_\Sigma = \frac{3}{2} \frac{M_\Sigma}{\lambda_{75}},
\eeq
where $\alpha, \beta, ...$ are the $SU(3)$ indices and $a,b,...$ are for $SU(2)$.
This provides different masses for different components of $\Sigma$ and splits the {\bf 75} dimensional multiplet.  

The second term of the Higgs superpotential Eq.~\eqref{eq:WH} is given by
\beqn
W_2 \,=\, g_H H \Sigma \Theta + g_{\overline H} {\overline H} \Sigma {\overline \Theta} + g_X \Theta {\overline \Theta} X \,.
\eeqn
Since ${\bf 50}$ does not contain a colour singlet $SU(2)$ doublet, the second and third terms cannot give the mass to the MSSM Higgs multiplets.
On the other hand, the colour triplet Higgses get masses from the two VEVs, $\langle \Sigma \rangle$ and $\langle X \rangle$, given by
\beqn
\hat m_{H_C} \,=\,
\begin{pmatrix}
0 & 4 \sqrt{3} g_H V_\Sigma \\
4 \sqrt{3} g_{\overline H} V_\Sigma & M_{\Theta}
\end{pmatrix},
\eeqn
with
\beq
M_{\Theta} = g_X \langle X \rangle \,.
\eeq
By diagonalising this matrix one finds the two mass eigenvalues
\beqn
M_{H_{C1(2)}} = \frac{M_{\Theta}}{2} \mp \frac{1}{2} \
\sqrt{ M_{\Theta}^2 + 192 g_H g_{\overline H} V^2_\Sigma} \,.
\eeqn

The GUT mass spectrum and the contribution to the $\beta$-function coefficients are given in Table~\ref{tb:mpsu5_2}.
\begin{table}[h!]
\begin{center}
\begin{tabular}{|c|c|c|c|}
\hline
$(b_1, b_2, b_3)$  &  ~~~mass~~~ & $(U(1) \times $SU(2)$ \times $SU(3)$)$ & $SU(5)$  \\
\hline
$(-10, -6, -4)$ & $M_V$ & $(-\frac{5}{6}, {\bf 2}, {\bf 3}),~ (\frac{5}{6}, {\bf 2}, {\bf \overline 3})$ &
\\ \hline
$(\frac{2}{5}, 0, 1)$ & $ M_{H_{C1}} $ & $(-\frac{1}{3}, {\bf 1}, {\bf 3}),~(\frac{1}{3}, {\bf 1}, {\bf \overline 3})$ & 
\\ \hline
$(\frac{2}{5}, 0, 1)$ & $ M_{H_{C2}} $ & $(-\frac{1}{3}, {\bf 1}, {\bf 3}),~(\frac{1}{3}, {\bf 1}, {\bf \overline 3})$ & 
\\ \hline
$(0, 16, 9)$ & $ M_\Sigma $ & $(0, {\bf 3}, {\bf 8})$ & 
\\ \hline
$(10, 0, 1)$ & $ \frac{4}{5} M_\Sigma $ & $(\frac{5}{3}, {\bf 1}, {\bf 3}),~(-\frac{5}{3}, {\bf 1}, {\bf \overline 3})$ & 
\\ \hline
$(10, 6, 10)$ & $ \frac{2}{5} M_\Sigma $ & $(\frac{5}{6}, {\bf 2}, {\bf 6}),~(-\frac{5}{6}, {\bf 2}, {\bf \overline 6})$ & 
\\ \hline
$(0, 0, 0)$ & $ \frac{2}{5} M_\Sigma $ & $(0, {\bf 1}, {\bf 1})$ & 
\\ \hline
$(0, 0, 3)$ & $ \frac{1}{5} M_\Sigma $ & $(0, {\bf 1}, {\bf 8})$ & 
\\ \hline
$(\frac{175}{5},35,34)$ & $ M_\Theta $ &  & $\{{\bf 50},~{\bf \overline{50}}\} - \{(-\frac{1}{3}, {\bf 1}, {\bf 3}),~(\frac{1}{3}, {\bf 1}, {\bf \overline 3})\}$ 
\\ \hline
\end{tabular}
\caption{The GUT mass spectrum and the contribution to the $\beta$-function coefficients.
    \label{tb:mpsu5_2}
}
\end{center}
\end{table}
Substituting these values into Eq.~\eqref{eq:formula_g} we find
\beqn
\Omega_G &=& 0.7291 \,10^{-4} \cdot M_{H_C}^{\frac{36}{19}} ( M_V^2 M_\Sigma )^{-\frac{6}{19}} M_{\Theta}^{-\frac{18}{19}}, 
\label{eq:Og_hagiwara}
\\
T_G &=& 0.1683 \cdot M_{H_C}^{\frac{8}{19}} (M_V^2 M_\Sigma )^{\frac{5}{19}} M_{\Theta}^{-\frac{4}{19}},
\label{eq:Tg_hagiwara}
\\
C_G &=& -46.972 
- 30.842 \ln \frac{\Lambda}{M_\Theta} 
- 20.947 \ln \frac{\Lambda}{M_\Sigma}  
- 8.316 \ln \frac{\Lambda}{M_{H_C}}  
+ 8.1053 \ln \frac{\Lambda}{M_V} , 
\label{eq:Cg_hagiwara}
\eeqn
with 
\beq
M_{H_C} = \sqrt{M_{H_{C1}} M_{H_{C2}}} \,=\, 48 g_H g_{\overline H} V_\Sigma \,.
\eeq
The numerical factors in Eqs.~\eqref{eq:Og_hagiwara} and \eqref{eq:Tg_hagiwara}
and the first term of Eq.~\eqref{eq:Cg_hagiwara} come from the fractional numbers
appearing 
in the masses of $\Sigma$ components in Table~\ref{tb:mpsu5_2}.
Note that all terms in $C_G$, except for the last one, are large and negative.
This  drives the gauge couplings into a non-perturbative regime 
much before the unification scale, as shown below. 

The condition for the gauge coupling unification Eq.~\eqref{eq:GCUcond} can be recast into
\beqn
\frac{M_{H_C}^2}{M_\Theta} &=&  5.47831 \, 10^{5} \cdot M_G^* \, \Omega_S \left( \frac{T_S}{M_S^*} \right)^{\frac{5}{6}} \,, 
\label{eq:lareg_ratio}
\\
(M_V^2 M_\Sigma)^{\frac{1}{3}} &=& 0.71554 \cdot M_G^* \, \Omega_S \left( \frac{T_S}{M_S^*} \right)^{-\frac{2}{9}} \,. 
\label{eq:MVS}
\eeqn 
It is evident 
from Eq.~\eqref{eq:lareg_ratio} that for reasonable SUSY spectra $M_\Theta$ has to be much smaller than $M_{H_C}$.
However, such configurations of the  GUT masses 
are incompatible with  perturbative gauge couplings unification.
As can be seen in Eq.~\eqref{eq:unifiedcoupling}, the contribution from the GUT threshold 
to the inverse of unified gauge coupling, $\alpha^{-1}(\Lambda)$, is given by $C_G/(2 \pi)$, which can be written by
\beqn
 \frac{C_G}{2 \pi} 
&=& -7.476 - 4.909 \ln \frac{M^2_{H_C}}{M_\Theta} 
- 11.14 \ln \frac{\Lambda}{M_{H_C}} 
- 3.334 \ln \frac{\Lambda}{M_{\Sigma}} 
+ 1.290 \ln \frac{\Lambda}{M_V} 
\nonumber \\ 
&=& -72.34 + 4.909 \ln \frac{\Lambda}{S} 
- 11.14 \ln \frac{\Lambda}{M_{H_C}} 
- 3.334 \ln \frac{\Lambda}{M_{\Sigma}} 
+ 1.290 \ln \frac{\Lambda}{M_V} ,
\label{eq:daG}
\eeqn 
with $S = M_G^* \, \Omega_S \left( T_S/M_S^* \right)^{\frac{5}{6}}$.
Here we used Eq.~\eqref{eq:lareg_ratio} in the second equality.
One can see that the first term is negative and much larger in magnitude than 
the constant term $\alpha^{*-1}_{G} = 25.5$ in Eq.~\eqref{eq:unifiedcoupling}.
Since $M_V$ cannot be taken much smaller than $10^{16}$ GeV to satisfy the $D=6$ proton decay constraint,
the last term of Eq.~\eqref{eq:daG} cannot be large.
Therefore, we conclude that it is not possible in this model to achieve the perturbative gauge coupling unification 
in phenomenologically allowed parameter region, unless the SUSY contribution $\frac{C_S}{2\pi}$ 
is positive and very large and/or $S \ll \Lambda$.
We do not consider such a possibility since it would require extreme mass hierarchies in the MSSM spectrum.

\subsection{Hisano-Moroi-Tobe-Yanagida Model}
\label{sec:yanagida}

A solution to the problem found in the Hagiwara-Yamada model was proposed by Hisano, Moroi, Tobe and Yanagida \cite{Hisano:1994fn}.
The main idea is to implement a structure to suppress the $D=5$ proton decay
and to make the $\Theta$ fields very heavy so that the second term in Eq.~\eqref{eq:Cg_hagiwara} is made small.
In their model new fields, distinguished with primes, are introduced in the $H, \overline H, \Theta, \overline \Theta$
sectors with appropriate $U(1)_X$ charges: $H'(2), \overline H'(-1), \Theta'(1), \overline \Theta'(-2)$.
Notice that now $\Theta$ fields can have tree-level mass terms with primed fields.
The superpotential for the primed fields can be written as
\beqn
W^\prime = g'_{\overline H}\overline H' \Sigma \Theta' + g'_{ H}  H' \Sigma \overline \Theta'
+ M_1 \overline \Theta \Theta' + M_2 \overline \Theta' \Theta
+ g_X \overline H' H' X \,.
\eeqn

We assume $M_1, M_2 \gg \langle \Sigma \rangle, \langle X \rangle$ so that
the mass splitting within the $\Theta$ multiplets can be neglected,
and the effective superpotential after integrating out the $\Theta$ fields can be used at the energy scale of the coupling unification. 
Substituting the VEVs of $\Sigma$ and integrating out the $\Theta$ fields, one finds the effective colour triplet Higgs mass
terms 
\beq
M_{H_{C_1}}\overline H_C \overline H'_C + M_{H_{C_2}}\overline H'_C \overline H_C \,,
\eeq
with
$M_{H_{C_1}} \simeq 48 g_H g'_{\overline H} V_\Sigma^2/M$
and $M_{H_{C_2}} \simeq 48 g'_H g_{\overline H} V_\Sigma^2/M$,
where $M_1 = M_2 = M$ is taken for simplicity. 
The last term of $W^\prime$ introduces the mass to the  doublet and triplet fields 
in $H', \overline H'$
\beq
M_f \big( \overline H'_f H'_f + \overline H'_C H'_C \big)\,,
\eeq
with $M_{H_f} \equiv g_X \langle X \rangle$.

Due to the $U(1)_X$ symmetry, $H'$ and $\overline H'$ cannot have Yukawa interactions with matter fields.
Since there is no direct mass term between $H_C$ and $\overline H_C$,
the propagators of $H_C$ and $\overline H_C$ cannot be connected  
in the $D=5$ proton decay operator by themselves.
The leading contribution comes via the mixing 
between the primed and unprimed triplet Higgs fields together with 
the direct mass term 
$M_{H_f} \overline H'_C H'_C$.
Therefore, the $D=5$ proton decay operator receives an extra suppression 
$M_{H_f} / M_{H_C}$ compared to the previous model with $M_{H_C} = \sqrt{M_{H_{C1}} M_{H_{C2}}}$.

The change to our formulae for $\Omega_G, T_G, C_G$ from the previous model is as follows.
Now the entire $\Theta, \overline \Theta, \Theta', \overline \Theta'$ multiplets 
(including triplet components) are decoupled and absent at the scale of the coupling unification.
Instead, new pairs of triplet and doublet (coming from $H'$ and $\overline H'$)
are present and contributing to the GUT threshold correction.
For $\Omega_G$ and $T_G$, we found the same expressions as Eqs.~\eqref{eq:Og_hagiwara} and \eqref{eq:Tg_hagiwara}
by replacing $M_{\Theta} \to M_{H_f}$
and $M_{H_C} = \sqrt{M_{H_{C1}} M_{H_{C2}}} \,=\, 48 \sqrt{g_H g_{\overline H} g'_H g'_{\overline H}} \, V^2_\Sigma / M$.
Because of this, the conditions of the gauge coupling unification, 
Eqs.~\eqref{eq:lareg_ratio} and \eqref{eq:MVS}, are also unchanged
up to the replacement $M_{\Theta} \to M_{H_f}$ in the former, which reads 
\beq
\frac{M_{H_C}^2}{M_{H_f}} =  5.47831 \, 10^{5} \cdot M_G^* \, \Omega_S \left( \frac{T_S}{M_S^*} \right)^{\frac{5}{6}} \,.
\label{eq:lareg_ratio2}
\eeq
For $\frac{C_G}{2 \pi}$, we have
\beqn
\frac{C_G}{2 \pi} &=& -7.476 
+ 0.502 \ln \frac{\Lambda}{M_{H_f}}
- 1.323 \ln \frac{\Lambda}{M_{H_C}} 
- 3.333 \ln \frac{\Lambda}{M_\Sigma} 
+ 1.290 \ln \frac{\Lambda}{M_V}
\nonumber \\
&=&
-0.834
-0.502 \ln \frac{\Lambda}{S}
-0.318 \ln \frac{\Lambda}{M_{H_C}} 
- 3.333 \ln \frac{\Lambda}{M_\Sigma} 
+ 1.290 \ln \frac{\Lambda}{M_V} \,,
\label{eq:Cg_yanagida}
\eeqn
where we have used Eq.~\eqref{eq:lareg_ratio2} in the second equality.
%
\begin{figure}[!t!]
\begin{center}
\includegraphics[width=9.cm]{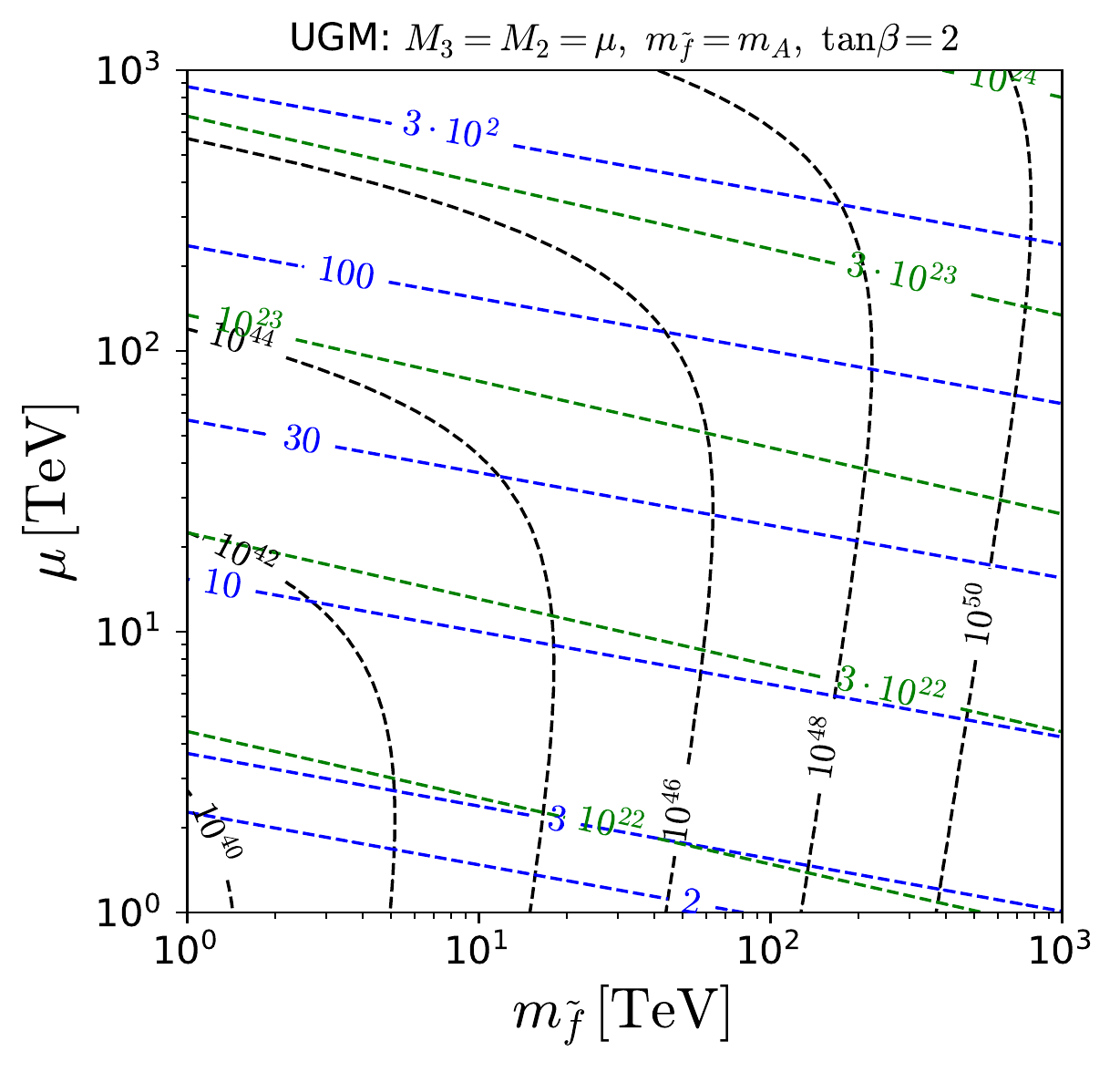}
\vspace{-2mm}
\caption{\label{fig:yan}
\small The low energy SUSY parameter space ($m_{\tilde f}$ versus $\mu$ plane)
for the UGM scenario with $M_3 = M_2 = \mu, m_A = m_{\tilde f}$ and $\tan\beta = 2$.
The dashed black, green and blue contours represent the values of $\tau(p \to K^+ \bar \nu)$/yrs,
$M_{H_C}$/GeV and $T_S$/TeV, respectively, calculated in the missing partner $SU(5)$ model in Ref.~\cite{Hisano:1994fn}.
}
\end{center}
\end{figure}
Contrary to Eq.~\eqref{eq:daG},
we see that the GUT contribution to the unified coupling is much smaller than the leading constant term $\alpha_G^* = 25.5$.
The perturbative coupling unification can therefore be easily achieved with ordinary SUSY spectra.   
The $D=5$ proton decay  depends on the ``effective triplet''  mass given by the ratio $M^2_{H_C}/M_{H_f}$  and, as shown in
Fig.~\ref{fig:yan}, is far beyond the experimental reach (in a foreseeable future)  for a realistic SUSY spectrum.  
The $D=6$ dependent decay channel depends on the $\Sigma$ mass, which is a free parameter of the model.
We report the ranges of the $(M_V^2 M_\Sigma)^{\frac{1}{3}}$ and $\alpha^{-1}(\Lambda)$ (computed assuming $M_V = M_\Sigma$)
obtained in the scan of the ($m_{\tilde f}$, $\mu$) plane in Fig.~\ref{fig:yan}:
\beq
3.21 \cdot 10^{15}\,{\rm GeV} ~< (M_V^2 M_{\Sigma})^{\frac{1}{3}} <~ 1.49 \cdot 10^{16}\,{\rm GeV} \,,
\eeq
\beq
39.6 ~<~ \alpha^{-1}(\Lambda) ~<~ 47.9~,
\eeq
where the nominal value of $\alpha_s(m_Z)$ was used.


\section{ {$SU(5)$ Orbifold  SUSY GUT}}
\label{sec:obf}


In this Section we  study the implication of  gauge coupling unification in the 5d $SU(5)$ orbifold SUSY  model presented in Ref.~\cite{Hall:2001pg}.
In this model, space-time is assumed to be the cross product of ordinary 4d Minkowski spacetime with a
$S^1 / (Z_2 \times Z_2^\prime)$ orbifold.
This orbifold has two fixed points $O$ and $O^\prime$ and one can assign to the fields
two independent $Z_2$ charges, $P$ and $P^\prime$,
corresponding to the reflection symmetries, $y \to -y$, of the $S^1$ coordinate $y$, 
centred around $O$ and $O^\prime$, respectively.
Since the spinor representation in the 5d spacetime has 8 real components, supersymmetry in the bulk is doubled 
compared to those in the 4d spacetime. 
As a consequence, the 5d vector multiplet ${\cal V}(V, \Sigma)$ contains 
a 4d vector multiplet $V$ and a 4d chiral multiplet $\Sigma$ in the adjoint representation. 
The Higgs fields are introduced in the bulk and embedded in the hypermultiplets, 
${\cal H}(H({\bf 5}), H^c({\overline {\bf 5}}))$ and ${\overline {\cal H}}(\overline H(\overline {\bf 5}), \overline H^c({\bf 5}))$, 
containing two 4d chiral multiplets in a vector like manner; 
\{$H$, $H^c\} \supset \{(H_F, H_C), (H^c_F, H^c_C)$\}
and $\{\overline H, \overline H^c \} \supset \{ (H_{\overline F}, H_{\overline C}), (H^c_{\overline F}, H^c_{\overline C})\}$, respectively, where the subscript $F$ ($C$) represents the doublet (triplet) component.
To break the $SU(5)$ into $G_{\rm SM}$, the orbifold parities are assigned so that only the 4d gauge multiplets corresponding to the
$G_{\rm SM}$ generators have zero modes.
More specifically, $V^a(+, +)$ and $V^{\hat a}(+, -)$ are taken for $(P, P^\prime)$, where
$a$ ($\hat a$) corresponds to the unbroken (broken) generators.
This parity assignment implies that the $SU(5)$ is broken at a 3-brane at $O^\prime$ but 
unbroken at the other brane at $O$.
In order to preserve the success of the charge quantization and assignment in the matter sector in the 4d GUTs,
the matter fields $F_i = (D^c, L)_i$ and $T_i = (Q, U^c, E^c)_i$, ($i=1,2,3$), are placed at the $SU(5)$ symmetric 3-brane at $O$.

\begin{table}[h!]
\centering
\begin{tabular}{|c|c|c|c|c|}
\hline
$KK$ mode  & mass & $(P, P^\prime)$ & ~~~~~~4d fields~~~~~~ & $\sum (b_1, b_2, b_3)$ \\ \hline \hline
zero     & 0    & $(+,+)$         & $V^a, ~H_F,~ H_{\overline F}$     &  \\ \hline \hline
\multicolumn{1}{ |c  }{\multirow{2}{*}{even} } &
\multicolumn{1}{ |c|  }{\multirow{2}{*}{$(2n+2)/R$} } & $(+,+)$ & $V^a,~ H_F,~ H_{\overline F}$ & 
{\multirow{2}{*}{$(\frac{6}{5}, -2, -6)$} } \\ \cline{3-4}
\multicolumn{1}{ |c  }{} & \multicolumn{1}{ |c|  }{} & $(-,-)$ 
& $\Sigma^a,~ H^c_F,~ H^c_{\overline F}$ & \multicolumn{1}{ c|  }{} \\ \hline \hline
\multicolumn{1}{ |c  }{\multirow{2}{*}{odd} } &
\multicolumn{1}{ |c|  }{\multirow{2}{*}{$(2n+1)/R$} } & $(+,-)$ & $V^{\hat a},~ H_C,~ H_{\overline C}$ &
\multirow{2}{*}{$(-\frac{46}{5}, -6, -2)$} \\ \cline{3-4}
\multicolumn{1}{ |c  }{} & \multicolumn{1}{ |c|  }{} & $(-,+)$ & $\Sigma^{\hat a},~ H^c_C,~ H^c_{\overline C}$ & 
\multicolumn{1}{ c|  }{} \\ \hline
\end{tabular}
\caption{\small The orbifold parity assignment and contributions to the $\beta$-function coefficients
from the bulk fields. $R$ is the compactification radius.
\label{tb:obf}}
\end{table}

In Table~\ref{tb:obf} we show 
the complete orbifold parity assignment for the bulk fields,
in addition to the total contribution to $\beta$-function coefficients 
from the even and odd $KK$-excitations.
In the table, we separate out the zero mode, since they are included in the MSSM.
With this charge assignment, the doublet-triplet splitting problem is elegantly solved 
because only two doublet Higgs fields can have zero modes.

The structure to suppress the $D=5$ proton decay discussed in subsection \ref{sec:yanagida}
is automatically implemented in this model.
This is because the $KK$-mass is generated only amongst the components residing in the same 5d multiplet; ~e.g.~ 
$W \supset \frac{(2n+1)}{R} \big[ H^{(2n+1)}_C H_C^{c(2n+1)} + H^{(2n+1)}_{\overline C} H_{\overline C}^{c(2n+1)} \big]$,
whereas there are no direct mass terms connecting two fields from different 5d multiplets, such as $H^{(2n+1)}_C H_{\overline C}^{(2n+1)}$ and $H^{c(2n+1)}_C H_{\overline C}^{c(2n+1)}$.
This can be also understood in terms of a $U(1)_R$ symmetry of this model
with the following charge assignment; $H(0), H^c(2), \overline H(0), \overline H^c(2), F_i(1), T_i(1)$.
One can see that due to this $U(1)_R$ symmetry only $H(0) \supset H_C$ and $\overline H(0) \supset H_{\overline C}$
can have Yukawa couplings to the matter fields. 
Since $H_C$ and $H_{\overline C}$ do not couple via a mass term, the $D=5$ proton decay operator is not generated.
In other words $D=5$ proton decay is forbidden by the $U(1)_R$ symmetry.


In calculating the threshold corrections we follow Ref.\cite{Hall:2001pg} and assume that the 5d theory is cut-off at a scale $\Lambda$ where the field theory is presumably incorporated in some more fundamental theory. 
There are two sources of  the GUT threshold corrections in this model.
One is from mass splitting among the $KK$-even and -odd mode multiplets, as shown in Table~\ref{tb:obf}.
This part can be treated with $T_G, \Omega_G$ and $C_G$ by the formula~\eqref{eq:formula_g}.
The other source is from the brane kinetic term at $O^\prime$.
This is because the 4d $SU(5)$ gauge symmetry is explicitly broken into $G_{\rm SM}$
at the $O^\prime$ brane
because there is no 4d gauge fields corresponding to the broken generators
at $O^\prime$. 
This means that one can introduce independent kinetic terms for the three MSSM gauge fields
with different gauge couplings.  
However, in general the bulk contribution to the 4d gauge coupling always dominates 
due to the spread of the wave function and the contribution from the brane kinetic term
is suppressed by the volume factor $2 \pi R \Lambda$ where $R$ is the radius of $S^1.$\footnote{For example, it has been estimated  in ref.~\cite{Hall:2001pg} that the contribution to the weak mixing angle from the brane kinetic terms is 
much less than 1\,\%  for $r\equiv R \Lambda = 4$}
In the following analysis, we neglect the contribution from the brane kinetic terms\footnote{Once the brane couplings are specified, their effect can be included into the analysis by adding appropriate constants in the set of Eq.~\eqref{eq:RGE} and in the first set of Eq.~\eqref{eq:ass}}.

 
Coming back to
the mass splitting among the $KK$ GUT multiplets due to the GUT breaking parity assignment. 
Neglecting a finite correction from the brane kinetic term, the following picture is expected \cite{Hall:2001pg}.
Evolving the gauge couplings from low energy to high energy, they approach each other in the MSSM.
After passing the compactification scale, $M_c = 1/R$, 
$KK$-modes appear and  the running changes.
In the $\mu > M_c$ regime, the running is slower but the gauge couplings 
continue to approach each other.
The three gauge couplings approximately meet at the cut-off scale, $\Lambda$,
where the 5d theory may be incorporated into a more fundamental theory.
Namely, the unification scale $\Lambda$ in Eq.~\eqref{eq:RGE} serves also as the cut-off scale in this model.

As before the gauge coupling unification condition can be expressed in terms of $T_G, \Omega_G$ and $C_G$.
From Eq.~\eqref{eq:formula_g} and Table~\ref{tb:obf}, one can see that the contributions to 
$\ln \Omega_G$, $\ln (T_G/\Lambda)$ and $C_G$ 
from all the odd $KK$ excitations with level $(2k + 1)$ takes the form 
\beq
c_o \ln \frac{2 k + 1}{r},
\eeq
where $r=\Lambda R$. Those from the even excitations with level $(2k + 2)$ are given by
\beq
c_e \ln \frac{2 k + 2}{r},
\eeq
where $c_o = -c_e = \frac{24}{19}$ for $\ln \Omega_G$,
$c_o = -c_e = \frac{18}{19}$ for $\ln (T_G/\Lambda)$
and 
$c_o = \frac{4}{19}$, $c_e = -\frac{156}{19}$ for $C_G$.
The index $k$ runs from 0 to $k^{o/e}_{\rm max}$ with 
$2 k^o_{\max} + 1 \leq r$ for odd and $2 k^e_{\max} + 2 \leq r$ for even excitations.
Summing over $k$ up to $k^{o/d}_{\rm max}$, we arrive at the expression 
\beqn
\Omega_G &=& 
\left[ \frac{ \prod_{k}^{k^o_{\rm max}} (2 k + 1) }{ \prod_k^{k^e_{\rm max}} (2 k + 2) } 
\Big( \frac{1}{r} \Big)^{k^o_{\rm max} - k^e_{\rm max} } 
\right]^{\frac{24}{19}},
\nonumber \\
\frac{T_G}{\Lambda} &=& 
\left[ \frac{ \prod_{k}^{k^o_{\rm max}} (2 k + 1) }{ \prod_k^{k^e_{\rm max}} (2 k + 2) } 
\Big( \frac{1}{r} \Big)^{k^o_{\rm max} - k^e_{\rm max} } 
\right]^{\frac{18}{19}},
\nonumber \\
C_G &=& \frac{4}{19} \ln \left[ \prod_{k=0}^{k^o_{\rm max}} \frac{2 k + 1}{r} \right]
- \frac{156}{19} \ln \left[ \prod_{k=0}^{k^e_{\rm max}} \frac{2 k + 2}{r} \right] \,.
\label{eq:obf_OTG}
\eeqn
Unlike in the 4d GUT models studied in previous sections,
$T_G$ and $\Omega_G$ are dependent on $\Lambda$ (explicitly or implicitly through $r$.
See Appendix \ref{ap:obf} for the reason why this is the case.)
We note that $\Omega_G$ and $T_G/\Lambda$ are simply related by
\beq
\frac{T_G}{\Lambda} = \Omega_G^{\frac{3}{4}} \,.
\label{eq:TO_relation}
\eeq

As examples, the explicit forms of $\Omega_G$ and $T_G/\Lambda$ for the first few ranges of $r$ are given in Table~\ref{tb:obf_example}.
\begin{table}[t!]
  \centering
  \renewcommand{\arraystretch}{1.5} 
  \begin{tabular}{c||c|c|c|c|c} \hline
     & $1 < r \le 2$ & $2 < r \le 3$ & $3 < r \le 4$ & $4 < r \le 5$ & $\cdots$ \\ \hline
    $\Omega_G$ & 
    $\left[ \frac{1}{r} \right]^{\frac{24}{19}}$ & 
    $\left[ \frac{1}{2} \right]^{\frac{24}{19}}$ & 
    $\left[ \frac{1 \cdot 3}{2} \frac{1}{r} \right]^{\frac{24}{19}}$ &
    $\left[ \frac{1 \cdot 3}{2 \cdot 4} \right]^{\frac{24}{19}}$ & $\cdots$ \\
    $T_G/\Lambda$ & 
    $\left[ \frac{1}{r} \right]^{\frac{18}{19}}$ & 
    $\left[ \frac{1}{2} \right]^{\frac{18}{19}}$ & 
    $\left[ \frac{1 \cdot 3}{2} \frac{1}{r} \right]^{\frac{18}{19}}$ &
    $\left[ \frac{1 \cdot 3}{2 \cdot 4} \right]^{\frac{18}{19}}$ & $\cdots$ \\
    \hline
  \end{tabular}
  \caption{\small
  The explicit forms for $\Omega_G$ and $T_G/\Lambda$ for the first few ranges of $r = R \Lambda$.
  \label{tb:obf_example}}
  \vspace{5mm}  
\end{table}
It is worth stressing that in this model $\Omega_G$, $T_G/\Lambda$ and $C_G$
are constants for a given $r$.
The first equation, $T_S = M_S^* \Omega_G$, of the unification condition Eq.~\eqref{eq:GCUcond}
therefore places a non-trivial constraint amongst the low energy SUSY masses as
\beq
T_S = \Big[ M_3^{-28} M_2^{32}  \mu^{12} m_A^{3}  X_{T} \Big]^{\frac{1}{19}} \,=\, M_S^* \Omega_G \,,
\label{eq:obf_constraint}
\eeq
where Eq.~\eqref{eq:formula_s} was used.
As examples, $T_S$ =(1612, 1276, 1031) GeV for $r=1.5$ and (779, 617, 498) GeV for $r=4$
for $\alpha_s(m_Z) = (0.1175, 0.1183, 0.1191)$.

The second equation of the unification condition is equivalent to 
$T_G/ \Lambda = M_G^* \Omega_S / \Lambda$,
where the left-hand-side is constant and can be traded with $\Omega_G^{\frac{3}{4}}$ using Eq.~\eqref{eq:TO_relation}.
This allows for the determination of the cut-off scale as a function of SUSY masses through $\Omega_S$ giving
\beqn
\Lambda &=& M_G^*  \Omega_G^{-\frac{3}{4}} \Omega_S
\nonumber \\
&=& 
M_G^* \Omega_G^{-\frac{31}{54}} M_s^{*\frac{19}{108}} \left( \frac{M_3}{M_2} \right)^{\frac{19}{216}} M_3^{-\frac{38}{216}} X_T^{-\frac{1}{108}} X_\Omega^{\frac{1}{288}}, 
\label{eq:obf_lambda}
\eeqn
where the second equality is obtained 
by computing $\Omega_S$ under the constraint of Eq.~\eqref{eq:obf_constraint}.
In this process we have eliminated the combination $(\mu^4 m_A)^{\frac{1}{5}}$.

Since $\Lambda$ is related to the mass of the first $KK$ $X,Y$ bosons by
$M_c=\Lambda/r$, for a given $r$ we can compute the $D=6$ proton decay, $p \to \pi^0 e^+$,
from the low energy SUSY spectrum .
\begin{figure}[!t!]
\begin{center}
\includegraphics[width=10.cm]{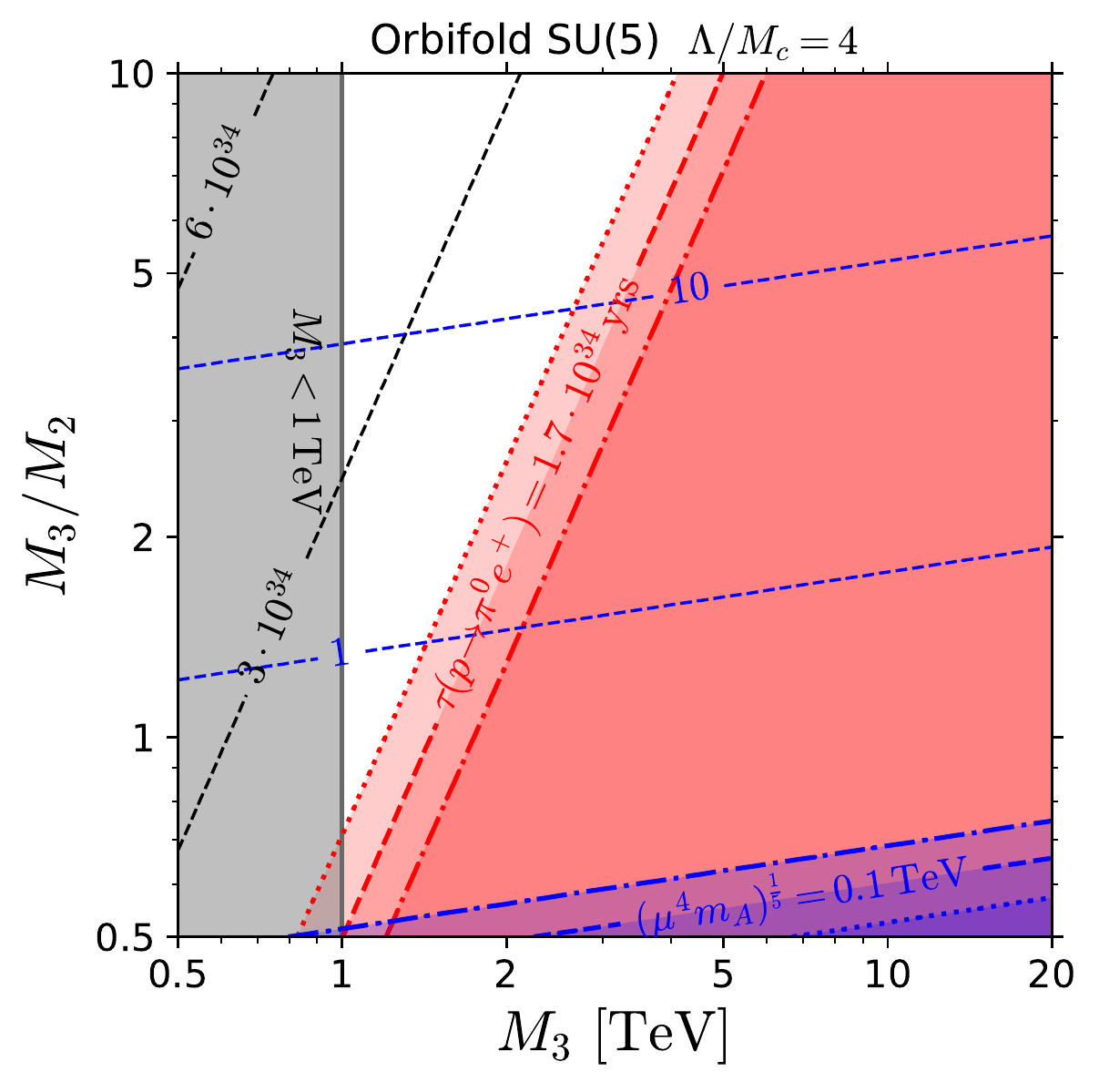}
\caption{\label{fig:obf}
\small The $M_3$ versus $M_3/M_2$ plane in the orbifold $SU(5)$ SUSY-GUT model \cite{Hall:2001pg} with $r = R \Lambda = 4$.
The grey shaded region shows a conservative limit on the gluino mass from the direct SUSY searches at the LHC.
The blue dashed lines are the contours of $(\mu^4 m_A)^{\frac{1}{5}}$
required by the gauge coupling unification.
The shaded blue region is phenomenologically excluded due to the presence of too light
charginos or non-SM Higgs bosons.
The black dashed lines are the contours of 
the $\tau(p \to \pi^0 e^+)$ predicted due to the condition of gauge coupling unification.
The shaded red region is excluded due to the present proton decay bound, 
$\tau(p \to \pi^0 e^+) < 1.7 \cdot 10^{34}$ yrs.
The black and blue dashed lines are obtained using the nominal value of $\alpha_s(m_Z)$,
whilst the dotted-dashed and dotted lines (red and blue)
are obtained using the upper and lower variation of $\alpha_s(m_Z)$ corresponding to the 1-$\sigma$ uncertainty.
}
\end{center}
\end{figure}
We show in Fig.~\ref{fig:obf} the ($M_3$, $M_3/M_2$) plane of the MSSM
for $\Lambda/M_c =1.5$  (left) and 4 (right).
For simplicity, we assume sfermion masses ($m_{\tilde f}$) are universal at low energy, which assures $X_T = X_\Omega = 1$.
We impose Eqs.~\eqref{eq:obf_constraint} and \eqref{eq:obf_lambda} 
so that the gauge couplings unify at $\Lambda$.
The former constraint allows one to determine $(\mu^4 m_A)^{\frac{1}{5}}$ at each point of the plane.
The dashed blue lines shows the contours of $(\mu^4 m_A)^{\frac{1}{5}}$ required for the unification using the nominal value of $\alpha_s$.
The region with $(\mu^4 m_A)^{\frac{1}{5}} < 100$ GeV marked by blue
predicts an unacceptably light chargino or non-SM Higgs bosons because one of the $\mu$ and $m_A$ (or both) is less than 100 GeV.
The dotted-dashed and dotted blue lines correspond to the same contour $(\mu^4 m_A)^{\frac{1}{5}} = 100$ GeV
but obtained from upper and lower 1-$\sigma$ variations of $\alpha_s(m_Z)$, respectively.

Light gluinos are strongly constrained 
by the null result of SUSY searches at the LHC.
We take the most conservative bound on the gluino mass \cite{Aaboud:2017vwy,Sirunyan:2018vjp,CMS-PAS-SUS-16-017}
and mark the excluded region, $M_3 < 1$ TeV, by grey.  
This limit generally applies if the spectrum is compressed,
while more severe bound should be applied otherwise.
When  the mass difference 
between gluino and the lightest SUSY particle (LSP) is large,\footnote{The running 
of the gauge coupling is independent of the Bino mass, which is a gauge singlet.  
The lower bound of the LSP mass is therefore unconstrained, since the LSP can be Bino-like.}
the gluino mass is excluded up to $\sim 2.0 - 2.2$ TeV by the current data \cite{Aaboud:2017vwy,Sirunyan:2018vjp,CMS-PAS-SUS-16-017,ATLAS-CONF-2018-041}.

As discussed above, the mass of the $X,Y$ gauge bosons is determined 
at each point of the parameter plane. 
We evaluate the unified coupling using Eq.~\eqref{eq:unifiedcoupling} and $C_S$ and $C_G$,
assuming $m_{\tilde f} = M_3$ and $\mu = m_A$, but the dependency on the SUSY spectrum is very mild.
Then, the proton decay lifetime $\tau(p \to \pi^0 e^+)$ can be calculated.
The black dashed lines show the contours of the $\tau(p \to \pi^0 e^+)$ using the nominal value of $\alpha_s$.
The current proton decay limit, $\tau(p \to \pi^0 e^+) < 1.7 \cdot 10^{34}$ yrs,
excludes the region shaded by red.
The dotted-dashed and dotted red lines represent the contours of $\tau(p \to \pi^0 e^+) < 1.7 \cdot 10^{34}$ yrs
obtained from upper and lower 1-$\sigma$ variations of $\alpha_s(m_Z)$, respectively.
The ranges of $\Lambda$ and $\alpha^{-1}(\Lambda)$ found for $r=4$ in the parameter range of the right plot in  Fig.~\ref{fig:obf} are
\beq
1.58 \cdot 10^{16} \, {\rm GeV} < \Lambda < 4.19 \cdot 10^{16} \, {\rm GeV}\,,
\eeq
\beq
26.2 < \alpha(\Lambda) < 28.4 \,.
\eeq

As can be seen, this model is constrained strongly by the LHC and the proton decay measurement.
For $r > 4$, the constraint is even tighter.
This is because the proton lifetime is roughly proportional to $M_c^4 = (\Lambda/ r)^4$. Although $\Lambda$ is larger for larger $r$ (see Eq.~\eqref{eq:obf_lambda} and the dependence of $\Omega_G$ on $r$) this effect is very mild as compared to the suppression by the fourth power of $r$ in the expression for $M_c$.  It also follows from Eq.~\eqref{eq:obf_lambda} that for fixed values of $r$  and the ratio $M_3/M_2$ proton life time is shorter for larger values of $M_3$ (the value of $\Lambda$ is then decreasing with increasing $M_3$). The upper bound on  $M_3$ following from the present limit on proton lifetime is larger for smaller value of $r$ and larger value of the ratio $M_3/M_2$, as both increase the value of $\Lambda$ (see Eq.~\eqref{eq:obf_lambda}). 
It is clear that further improvement of the collider constraint as well as the proton lifetime limit
will cover the entire parameter space of this model.

\section{Discussions and conclusions}

In this paper, we have derived analytic expressions
for the condition of gauge coupling unification (Eq.~\eqref{eq:GCUcond}) 
and the unified gauge coupling (Eq.~\eqref{eq:unifiedcoupling})
in terms of the masses of SUSY and GUT particles.  
The formula is generic and applicable for any GUT models 
in which the SM gauge group is directly unified into a simple unified gauge group
at some high energy scale.
The unification condition, Eq.~\eqref{eq:GCUcond}, is expressed
in a form of two simultaneous equations, which are simple (and symmetric) relations 
between the four variables $T_S$, $\Omega_S$, $T_G$ and $\Omega_G$.
This is advantageous because, no matter  how complicated the GUT models are,
the condition can be written in terms of only  those four variables.

The first two variables ($T_S$ and $\Omega_S$) are  functions of superparticle masses 
and their explicit forms in the MSSM are given in Eq~\eqref{eq:OTC}.
The formula is derived from the RGE equation
including the  2-loop effect, while the threshold correction is
treated at 1-loop.
Therefore, the condition is insensitive to the mass of a particle that is singlet 
under the SM gauge group, such as the Bino.
Thus, the gauge coupling unification (GCU) condition alone cannot determine what the lightest SUSY particle is in the MSSM. 
Similarly, the MSSM formula Eq~\eqref{eq:OTC} is unchanged 
even for singlet extensions of the MSSM, such as next-to-MSSM (NMSSM).
For non-singlet extensions of the MSSM,
the corresponding formula can be found straightforwardly 
by solving the second set of linear equations in Eq.~\eqref{eq:ass}
with Eq.~\eqref{eq:s}.

The remaining two variables ($T_G$ and $\Omega_G$) are 
functions of the GUT masses.
The expressions in Eq.~\eqref{eq:formula_g}
are  generic for any GUT model with gauge coupling unification.
Despite their appearance, $\Omega_G$ and $T_G$ are independent of the unification scale $\Lambda$
for conventional 4d GUT models.
On the other hand, they may be dependent on $\Lambda$ in GUT models in higher dimensions,
and we have seen this is indeed the case in the orbifold $SU(5)$ model in Section \ref{sec:obf}.
A more detailed discussion of the $\Lambda$ dependence of the condition for the  GCU is given in
the Appendix.

The GCU condition and the unified gauge coupling are of course 
subject to the uncertainty in the measured value of the strong coupling constant at the weak scale, $\alpha_s(m_Z)$.
In our formalism this uncertainty is conveniently treated by 
understanding the effect of $\alpha_s(m_Z)$ on the three constants $M_S^*$, $M_G^*$ and $\alpha_G^*$ appearing in the analytic formula.
The dependence on the value of  $\alpha_s(m_Z)$ of  these constants is numerically parametrised in Eq.~\eqref{eq:as_effect}.

In Section \ref{sec:msu5}, Minimal $SU(5)$ SUSY GUT
is studied using our analytical formulae.
We have found that the GCU conditions  are re-expressed 
as the formula for the coloured Higgs mass (Eq.~\eqref{eq:MHCsu5}),
and that for $(M_V^2 M_\Sigma)$ (Eq.~\eqref{eq:MASSsu5}), 
as functions of the superparticle masses.
Using the former formula, one can predict the $D=5$ proton decay
mediated by the superparticles and the coloured higgsinos entirely
in terms of the MSSM mass spectrum.
We have shown in Fig.~\ref{fig:su5} the $D=5$ proton decay lifetime 
in several slices of SUSY parameter space.
The $D=5$ proton decay constraint is quite severe for the 
Minimal $SU(5)$ SUSY models.
However, relatively light SUSY spectrum ($M_3, M_2, \mu \sim {\cal O}(1)$ TeV)
is possible if sfermion masses are taken to be large, $m_{\tilde f} \gtrsim 30$ GeV.

Two missing partner $SU(5)$ models have been discussed in Section \ref{sec:missing_su5},
where the doublet triplet splitting is naturally solved. 
In Hagiwara-Yamada model \cite{Hagiwara:1992ys}, we have shown analytically that 
the GCU condition cannot be made consistent with the perturbativity of gauge couplings 
with reasonable SUSY spectra.
In Hisano-Moroi-Tobe-Yanagida model \cite{Hisano:1994fn}, 
this problem is solved by making the  $\Theta$ field very heavy.
In this model, the $D=5$ proton decay is very suppressed and 
the predicted values are beyond the next generation nucleon decay experiments.

The 5d orbifold $SU(5)$ SUSY GUT model \cite{Hall:2001pg} has also been studied in Section \ref{sec:obf}.
We have shown that the variables $T_G$, $\Omega_G$ and $C_G$ are effectively functions of 
the cut-off scale (unification scale), $\Lambda$, and $r = R \Lambda$, which determine the $KK$ spectrum. 
The GCU condition imposes a non-trivial constraint on the MSSM spectrum
for given $r$, and the cut-off scale is also determined.
In Fig.~\ref{fig:obf} we have shown the collider and $D=6$ proton decay constraint 
in the ($M_3/M_2$ vs $M_3$) parameter plane.
Since the $X,Y$ gauge boson mass (i.e.~the compactification scale) is proportional to $M_3^{-\frac{38}{216}}$ 
(see Eq.~\eqref{eq:obf_lambda}) the 
larger  the  gluino mass the  faster $D=6$ proton decay is predicted.
Combining the gluino mass bound from the collider search, we have  found a complementarity
between the collider and proton decay experiments in  testing this model.
It has been shown that the 5d orbifold $SU(5)$ SUSY GUT \cite{Hall:2001pg} 
is already very severely constrained by the LHC and the $D=6$ proton decay measurement.



\medskip
\section*{Acknowledgements}
The work of SP, KR and KS is partially supported by the Beethoven grants DEC-2016/23/G/ST2/04301.
The work of SP and KR are partially supported by the Harmonia grants DEC-2015/18/M/ST2/00054. 
The work of KS is partially supported by the National Science Centre, Poland, under research grants 2017/26/E/ST2/00135.
The work of KR is partially supported by the National Science Centre, Poland, under research grants 2015/19/D/ST2/03136.


\appendix

\section{Interpretations of threshold corrections}

In this appendix we compare two different formulations and interpretations of 
the 1-loop threshold correction from superheavy particles 
and discuss the condition of the GCU in each case.
We also comment on orbifold GUT models,
where the GCU may look accidental
from the 4d field theoretical point of view.

\begin{figure}[!t!]
\begin{center}
\includegraphics[width=8.cm]{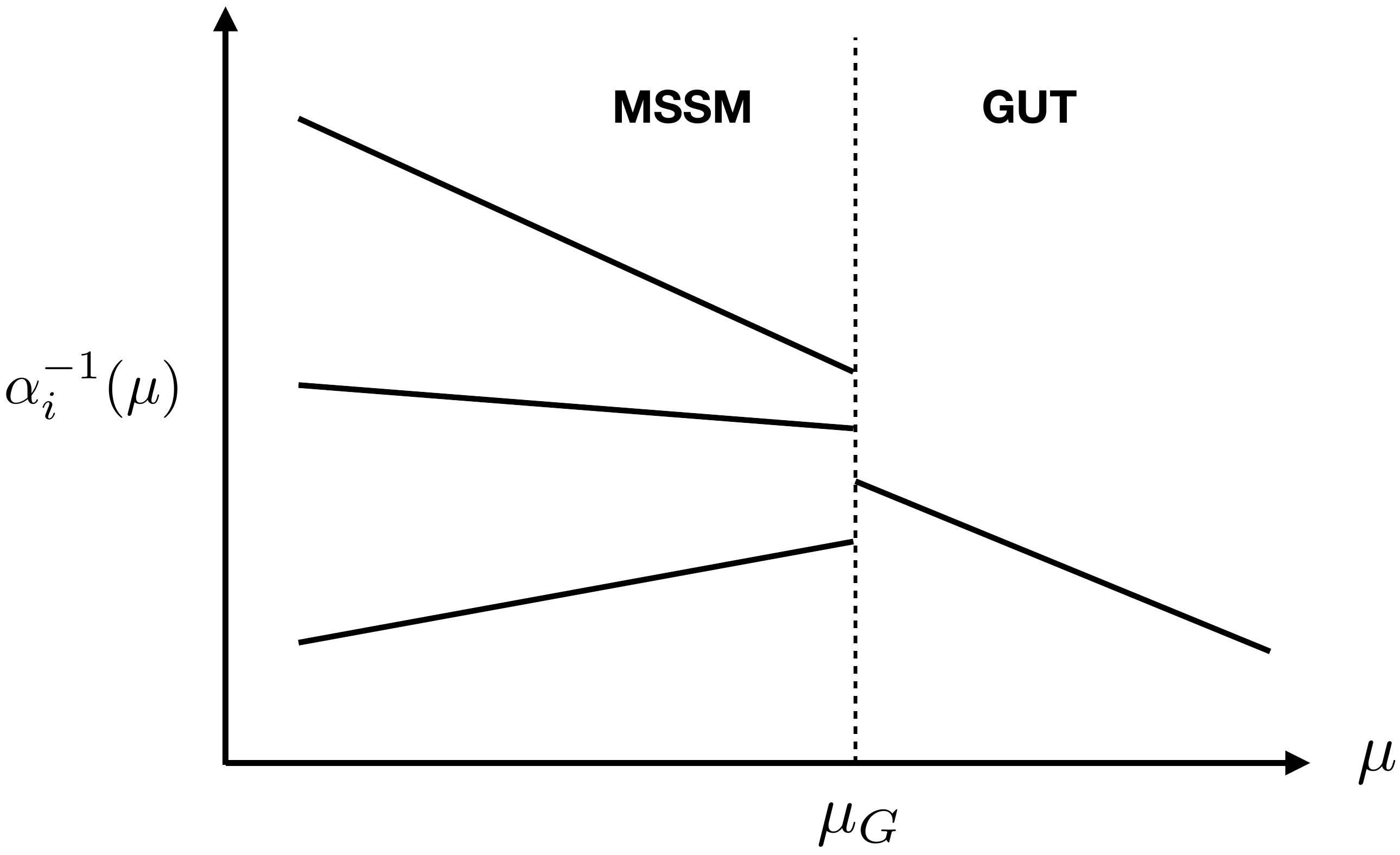}
\includegraphics[width=8.cm]{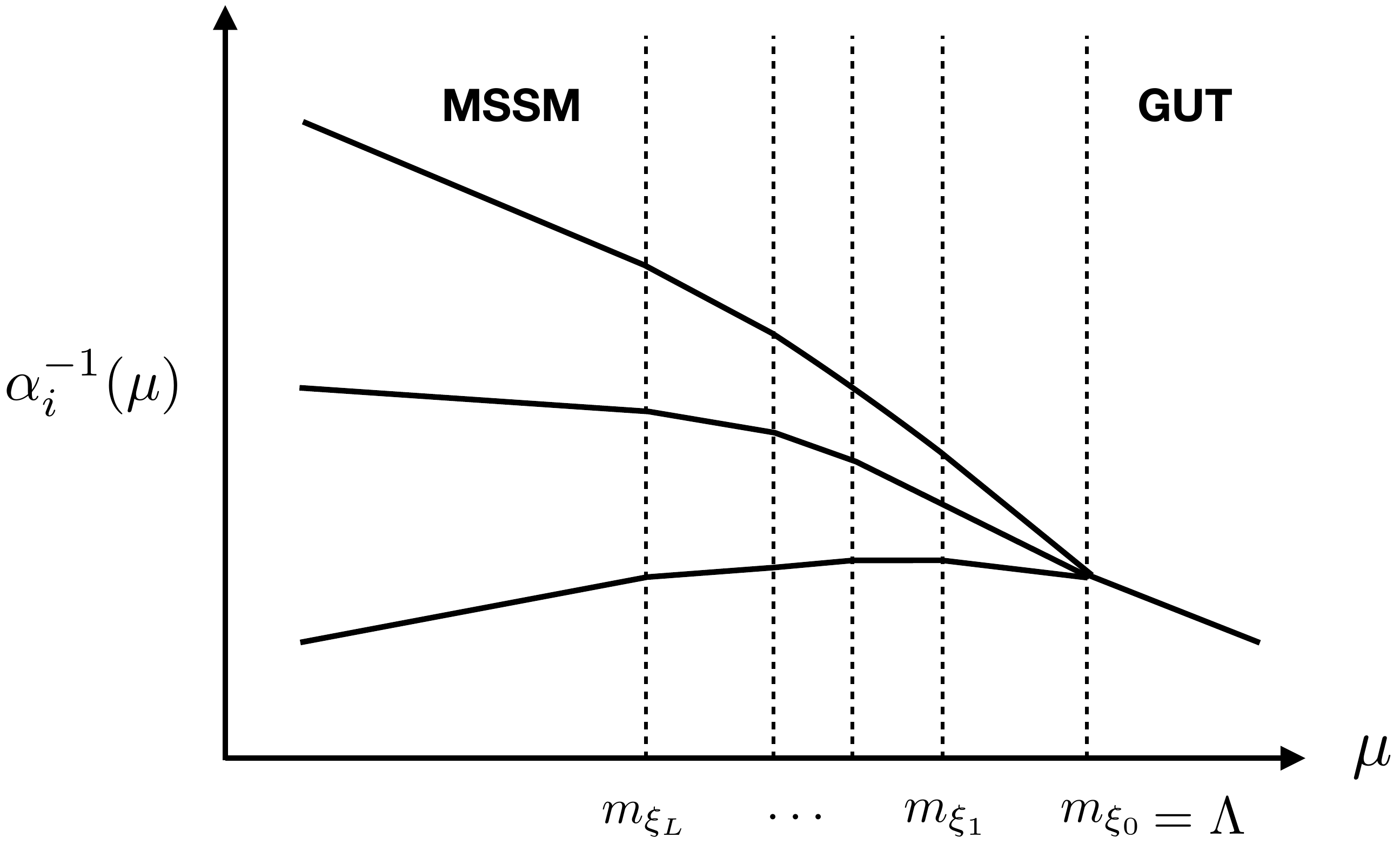}
\caption{\label{fig:evolutions}
\small The RGE evolution of the gauge couplings in two different pictures:
{\bf Left:} A single EFT matching (from GUT to MSSM) at $\mu_G$.
{\bf Right:} Sequential EFT matchings at the particle masses.
Here, the unification scale, $\Lambda$, is given by the mass of the heaviest particle, $m_{\xi_0}$,
that forms an incomplete GUT multiplet.
}
\end{center}
\end{figure}

\subsection{Mass independent unified gauge coupling}

One way to find the relation between the MSSM parameters and the GUT model
is to derive the MSSM as a low energy effective field theory (EFT) of the GUT model
simply by integrating out all the superheavy particles and expressing the result in terms of the single, mass independent, minimal subtraction coupling constant of the GUT group\cite{LLewellynSmith:1980mu}. The strategy is to stay away from thresholds where the scale dependence of couplings is complicated. This is done by computing the difference between the coupling at $\mu_G$  below the heavy masses to the scale $\mu_2$ well above the masses, analytically integrating out the heavy states, and avoiding the need numerically to integrate through the threshold. The result, equivalent to the analyses in \cite{Weinberg:1980wa,Hall:1980kf}, may be written as the boundary condition for the running coupling constants 
\beq
\frac{2 \pi}{\alpha(\mu_G)} = \frac{2 \pi}{\alpha_i(\mu_G)} + r_i(\mu_G) \,,
\label{eq:single_match}
\eeq
where $\alpha(\mu_G)$ on the left-hand-side is the unified gauge coupling of the GUT model
and $\alpha_i(\mu_G)$ on the right-hand-side are the MSSM gauge couplings  evaluated at $\mu_G$.
The leading-log expression of the threshold correction (neglecting a small finite correction arising when integrating out the gauge bosons)
is given by \cite{Weinberg:1980wa, LLewellynSmith:1980mu, Hall:1980kf}
\beqn
r_i(\mu_G) \,=\, \sum_\xi b^\xi_i \ln \left( \frac{m_\xi}{\mu_G} \right),
\eeqn
which is the same as Eq.~\eqref{eq:r} with a replacement $\Lambda \to \mu_G$.
The condition of the GCU should be understood as the $i$-independence of the right-hand-side of Eq.~\eqref{eq:single_match}.
At first glance, the condition seems dependent on the matching scale $\mu_G$.
However, the condition is independent of $\mu_G$
because the $\beta$-function coefficient of the GUT model, $b_G$, is related to the MSSM 
$\beta$-function coefficients by $b_G = b_i + \sum_\xi b_i^\xi$.
The result is illustrated 
in the left panel of Fig.~\ref{fig:evolutions}. 

\subsection{Sequential matchings}

Another way to find the GCU condition is to sequentially construct EFTs
and carry out EFT matching every time when 
the renormalization scale $\mu$ crosses a mass of superheavy particle
(see the right panel of Fig.~\ref{fig:evolutions}).
Let us label the superheavy particles as $\xi_0, \xi_1, \cdots, \xi_L$
with $m_{\xi_0} > m_{\xi_1} > \cdots > m_{\xi_L}$,
and assume that the theory is no longer symmetric under the unified gauge group
by integrating out $\xi_0$.
At the scale $\Lambda = m_{\xi_0}$, we have the following matching condition
\beq
\frac{2 \pi}{\alpha(\Lambda)} = \frac{2 \pi}{\alpha_i(m_{\xi_0})} \,.
\label{eq:first_match}
\eeq
The threshold correction arising from integrating out $\xi_0$ 
(as in Eq.~\eqref{eq:single_match}) is vanishing in this case,
because the matching scale is taken to be $m_{\xi_0}$ and the logarithm vanishes.

Below $\Lambda$, the theory is no longer symmetric under the unified gauge symmetry
due to the absence of $\xi_0$ and three gauge couplings evolves differently.
In particular, the $\beta$ function coefficient is changed 
from that of the original GUT model, $b_G$ ($i$-independent),
to $b_G - b_i^{\xi_0}$ subtracting the contribution from $\xi_0$. 
We run down the three gauge couplings with the new coefficients 
$b_G - b_i^{\xi_0}$ to the scale $\mu = m_{\xi_1}$.
The solution to the 1-loop RGE gives us
\beq
\frac{2 \pi}{\alpha_i(m_{\xi_0})}  = \frac{2 \pi}{\alpha_i(m_{\xi_1})} + (b_G - b_i^{\xi_0}) 
\ln \left( \frac{m_{\xi_1}}{m_{\xi_0}} \right)\,.
\eeq
By repeating the same procedure and run down the gauge couplings to $\mu = m_{\xi_2}$, we have
\beq
\frac{2 \pi}{\alpha_i(m_{\xi_1})}  = \frac{2 \pi}{\alpha_i(m_{\xi_2})} + (b_G - b_i^{\xi_0} - b_i^{\xi_1}) 
\ln \left( \frac{m_{\xi_2}}{m_{\xi_1}} \right)\,.
\eeq
Substituting this to the above equations leads to
\beqn
\frac{2 \pi}{\alpha(\Lambda)} = \frac{2 \pi}{\alpha_i(m_{\xi_2})} 
+ b_G \ln \left( \frac{m_{\xi_2}}{\Lambda} \right)
- b_i^{\xi_0} \ln \left( \frac{m_{\xi_2}}{m_{\xi_0}} \right)
- b_i^{\xi_1} \ln \left( \frac{m_{\xi_2}}{m_{\xi_1}} \right) \,.
\eeqn
Repeating the process until the renormalization scale smaller than the lightest superheavy particle mass,
we find the relation between the MSSM gauge coupling at $\mu < m_{\xi_L}$ and the unified gauge coupling 
at $\Lambda$ as 
\beq
\frac{2 \pi}{\alpha(\Lambda)} = \frac{2 \pi}{\alpha_i(\mu)} 
+ b_G \ln \left( \frac{\mu}{\Lambda} \right)
- \sum_\xi b_i^{\xi} \ln \left( \frac{\mu}{m_{\xi}} \right)\,.
\label{all_match}
\eeq
Since the unified gauge coupling evolves as
\beq
\frac{2 \pi}{\alpha(\Lambda)} = \frac{2 \pi}{\alpha_i(\mu_G)} 
+ b_G \ln \left( \frac{\mu_G}{\Lambda} \right) \,,
\eeq
in the GUT model, Eq.~\eqref{all_match} holds at arbitrary scale around $\mu_G \sim 10^{16}$\,GeV.
\beqn
\frac{2 \pi}{\alpha(\mu_G)} = \frac{2 \pi}{\alpha_i(\mu)} 
+ b_G \ln \left( \frac{\mu}{\mu_G} \right)
- \sum_\xi b_i^{\xi} \ln \left( \frac{\mu}{m_{\xi}} \right) \,.
\label{all_match_muG}
\eeqn
Here, the condition of GCU is understood as the 
$i$-independence of the sum of the first and third terms in the right-hand-side.
It is apparent that the condition depends neither on $\Lambda$ nor $\mu_G$ in this formalism.

It is straightforward to do the similar exercise but 
evolving gauge couplings in the opposite direction (from low energy to high energy).
We find in this case,
\beq
\frac{2 \pi}{\alpha(\mu_G)} = \frac{2 \pi}{\alpha_i(\mu)} 
+ b_i \ln \left( \frac{\mu}{\mu_G} \right)
+ \sum_\xi b_i^{\xi} \ln \left( \frac{m_{\xi}}{\mu_G} \right)\,.
\label{all_match_muG_2}
\eeq
Eq.~\eqref{all_match_muG} and \eqref{all_match_muG_2} are of course equivalent 
with the relation 
\beq
b_G = b_i + \sum_\xi b_i^{\xi}\,.
\eeq
The combination of the first and second terms in the right-hand-side of 
Eq.~\eqref{all_match_muG_2}
is nothing but the MSSM gauge couplings at $\mu_G$.
Thus, it reproduces the previous result Eq.~\eqref{eq:single_match}
obtained from the first approach, showing 
the equivalence between the two formalisms.

\subsection{A comment on orbifold GUTs}
\label{ap:obf}

In the above two subsections we have seen that the condition of GCU is 
independent of both the matching scale $\mu_G$ and the unification scale $\Lambda$, which appears in the second picture in Fig.~4,
provided the $\beta$-function coefficients of the MSSM is related to that of the GUT model
by $b_i = b_G + \sum_\xi b_i^\xi$.
In the orbifold GUT model, however, the unified gauge symmetry 
is never realised even at arbitrary high energies in the 4d space-time
and $b_G$ therefore does not exist.

Neglecting a finite correction from the brane kinetic term mentioned in the main text,
the following picture is expected \cite{Hall:2001pg}.
Renormalizing from low energy to high energy, the three gauge couplings of the MSSM approach each other.
After passing the compactification scale, $M_c$, 
$KK$-modes appear and the running changes.
In the $\mu > M_c$ regime the running is slower but the gauge couplings 
continue to approach each other.
At a scale, $\Lambda$, it is assumed that the three gauge couplings finally meet
and the 5d theory may be incorporated into a more fundamental theory.
Thus, the unification scale $\Lambda$ serves as the cut-off scale of the
5d theory.  

Since there is no unified gauge theory in the 4d space-time,
the 4d EFT matching between the GUT model and its low energy EFT does not make sense,
and we are forced to use the bottom-up RGE evolution and Eq.~\eqref{all_match_muG_2}
with $\mu_G = \Lambda$.
Since Eq.~\eqref{all_match_muG_2} cannot be related to Eq.~\eqref{all_match_muG},
the condition of GCU in orbifold GUTs does depend on $\Lambda$,
as can be seen in Section \ref{sec:obf}.

\addcontentsline{toc}{section}{References}
\bibliographystyle{JHEP}
\bibliography{ref}

\end{document}